\documentclass{article}
\usepackage[utf8]{inputenc}
\usepackage{amsmath,amssymb}
\usepackage{bm, bbm, blkarray, adjustbox}
\usepackage{amsfonts}
\usepackage{enumitem}
\usepackage[margin=1in]{geometry}
\usepackage{natbib}
\usepackage{hyperref}
\hypersetup{
  colorlinks = true,
  linkcolor = blue,
  urlcolor  = blue,
  citecolor = blue,
  anchorcolor = blue}
  
\usepackage{graphicx}
\usepackage{soul}
\usepackage{setspace}
\usepackage{listings}
\usepackage{color,soul}
\usepackage{authblk}
\usepackage{booktabs}
\usepackage{pdflscape}
\usepackage{algorithm,algorithmicx,algpseudocode}

\usepackage[normalem]{ulem}
\usepackage{subcaption}

\usepackage{tikz}
    \usetikzlibrary{arrows,positioning}

\newcommand\bovermat[2]{%
  \makebox[0pt][l]{$\smash{\overbrace{\phantom{%
    \begin{matrix}#2\end{matrix}}}^{\text{#1}}}$}#2}

\newlist{condenum}{enumerate}{1} 
\setlist[condenum]{label=\bfseries Condition \arabic*., ref=\arabic*, wide}

\title{Bayesian Cox model with graph-structured variable selection priors for multi-omics biomarker identification}

\author[1]{Tobias Østmo Hermansen}
\author[2,3]{Manuela Zucknick}
\author[2*]{Zhi Zhao}

\affil[1]{Department of Mathematics,
University of Oslo, Oslo, Norway}

\affil[2]{Oslo Centre for Biostatistics and
Epidemiology, Department of Biostatistics,
University of Oslo, Oslo, Norway}

\affil[3]{Oslo Centre for Biostatistics and
Epidemiology, Research Support Services, Oslo University Hospital, Oslo, Norway}

\affil[*]{\textit{email: zhi.zhao@medisin.uio.no}}

\begin{document}
\date{}

\maketitle

\begin{abstract}
An important goal in cancer research is the survival prognosis of a patient based on a minimal panel of genomic and molecular markers such as genes or proteins. 
Purely data-driven models without any biological knowledge can produce non-interpretable results. 
We propose a penalized semiparametric Bayesian Cox model with graph-structured selection priors for sparse identification of multi-omics features by making use of a biologically meaningful graph via a Markov random field (MRF) prior to capturing known relationships between multi-omics features. 
Since the fixed graph in the MRF prior is for the prior probability distribution, it is not a hard constraint to determine variable selection, so the proposed model can verify known information and has the potential to identify new and novel biomarkers for drawing new biological knowledge. 
Our simulation results show that the proposed Bayesian Cox model with graph-based prior knowledge results in more trustable and stable variable selection and non-inferior survival prediction, compared to methods modeling the covariates independently without any prior knowledge. 
The results also indicate that the performance of the proposed model is robust to a partially correct graph in the MRF prior, meaning that in a real setting where not all the true network information between covariates is known, the graph can still be useful. 
The proposed model is applied to the primary invasive breast cancer patients data in The Cancer Genome Atlas project.
\end{abstract}


\section{Introduction}

Personalized molecular medicine makes use of patient-specific genomic and molecular markers that are indicative of the development of disease processes to improve individualized diagnostic and therapeutic approaches. 
To capture the complex relationships between massive omics feature sets and the time to an event of interest (e.g. time to death or disease progression), Bayesian survival modeling provides a flexible framework. 
In the Bayesian framework, it is common to use shrinkage priors and variable selection indicators for modeling the effects of high-dimensional omics features, which can induce a minimal panel of identified omics features used as biomarkers to enhance patient management and personalized treatment strategies. 

Based on the commonly used Cox proportional hazards model (Cox model) \citep{Cox75}, Bayesian lasso-type Cox models \citep{Lee2011, Lee2015, Zucknick2015} were proposed for modeling high-dimensional input data, but they do not implement straightforward variable selection. 
Stochastic search variable selection (SSVS) is an attractive alternative approach to identify important covariates through latent indicator variables, in particular by using the popular independent spike-and-slab priors for variable selection \citep{George1993, Konrath2013, Treppmann2017}. 
It is well-known that genomic and molecular features do not work independently but generally interact with one another, and they often function together through multiple signaling routes or networks. 
\cite{Stingo2011} used a Markov random field (MRF) prior to incorporate gene dependence structures in a Bayesian parametric survival model for pathway and gene selection. 
\cite{Peterson2015} and \cite{Madjar2021} extended the model by \cite{Stingo2011} to infer the dependencies of omics features via graphical learning in a Bayesian parametric survival model and semiparametric Cox model, respectively. 
However, in the presence of multicollinearity (that is the usual case in high-dimensional multi-omics data), the resulting highly multi-modal spike-and-slab posteriors can be hostile to exploration, leading to unreliable and unstable variable selection \citep{Rokov2014, Rokov2018}. 
 
In general, it is challenging to take into account the complex structures in multi-omics data and to consider both technical and biological intricacies of the data and technologies \citep{Wissel2023}. 
\cite{Valous2024} discussed an integrative multi-omics data analysis using graph machine learning approaches, and suggested a robust integrative analysis of multi-omics data with network-based prior knowledge. 
\cite{Zhao2024jrssc} already showed the robustness of variable selection from multi-omics data through the spike-and-slab and MRF priors with a fixed informative graph prior, but for modeling continuous outcomes rather than time-to-event outcomes. 
In this article, we adapt the Bayesian Cox model in \cite{Madjar2021} by using a fixed informative graph in the MRF prior to identify covariates for survival prognosis and systematically investigated the impact of the MRF prior on model performance, both in terms of variable selection and prediction accuracy. 
Our proposed new model for inferring potential associations between multi-omics features and survival outcomes can leverage the network information known a priori to improve the model performance, and in particular to improve the stability of variable selection. 
Since the fixed graph in the MRF prior is for the prior probability distribution, it is not a hard constraint to determine variable selection, so the proposed model can verify known information and has the potential to identify new and novel biomarkers for drawing new biological knowledge. 

The rest of the article is organized as follows. 
In Section 2, we introduce the Bayesian Cox model and propose to use a fixed graph in an MRF prior for the latent variable selection indicators of the coefficients. 
Section 3 compares the performances of Bayesian Cox models with the MRF prior with an empty graph, completely correct graph and partially correct graphs in simulated data, to assess the sensitivity of model performances to incomplete or wrong prior information.
In Section 4, we analyze the survival and multi-omics data from breast cancer patients in The Cancer Genome Atlas (TCGA). 
In Section 5, we conclude the article with a discussion. 

\section{Methodology}
\label{sec:methods}

\subsection{Bayesian Cox model with informative MRF prior}
\label{sec:bayesCox_}

The Cox model \citep{Cox75} is to model the hazard function $h(t| \mathbf X_i)$ at time $t$ of a patient with $p$-row-vector covariates $\mathbf X_i=(X_{i1},...,X_{ip}) \in \mathbb R^p$ ($i=1,...,n$), and $\mathbf X \in \mathbb R^{n\times p}$ is the input data matrix of clinical and genomic covariates of $n$ patients. 
The Cox model is specified through a relative risk function, and assumes log-linearity of covariates and proportional hazards, i.e.
$$
h(t| \mathbf X_i) = h_0(t) \exp(\mathbf X_i\bm\beta),
$$
where the coefficients $\bm\beta =(\beta_1,...,\beta_p)^\top$ represent the effects of $p$ clinical and genomic covariates, and $h_0(t)$ is the baseline hazard rate. 
Let the right-censored data of $n$ patients be $\mathfrak D = \{ (T_i, \delta_i, \mathbf X_i): i=1,...,n \}$, where $T_i$ is the observed survival time and $\delta_i$ is the censoring indicator (i.e. $\delta_i=1$ means the occurrence of the event of interest by the end of the study and $\delta_i=0 $ otherwise). 
In the frequentist Cox model \citep{Cox75}, the baseline hazard $h_0(t)$ is left unspecified and the coefficients can be estimated by maximizing the partial likelihood 
$
\prod_{i=1}^n \left\{\frac{\exp(\mathbf{X}_i\bm\beta)}{\sum_{l\in \mathcal{R}_i }\exp(\mathbf{X}_l\bm\beta)}\right\}^{\delta_i},
$
where $\mathcal{R}_i$ is the risk set of patients at time $T_i$ that have not yet had the event and not yet been censored. 

In a Bayesian Cox model, the cumulative baseline hazards function $H_0(t)=\int_0^t h_0(s)\text{d} s$ can be modeled by a process prior, e.g. gamma process \citep{Kalbfleisch} or beta process \citep{Hjort1990}. 
The gamma process prior is popular and defined as
\begin{equation}\label{eq:H0_bayesCox}
H_0(t) \sim \mathcal{GP}(a_0 H^*(t), a_0),
\end{equation}
where $H^*(t)$ is an increasing function as the mean of the process with $H^*(0) = 0$, and $a_0 > 0$ as a weight or confidence parameter about the mean. 
The function $H^*(t)$ is often set to be a parametric function, e.g. the Weibull distribution $H^*(t) = \eta t^{\kappa}$, where $\eta>0$ and $\kappa>0$ are known hyperparameters. 
For practical purposes, it is often sufficient to use a discretized version of the gamma process. 
Consider a finite partition of time, $0 = c_0 < c_1 < \ldots < c_K$, such that $c_K > \max_{1\le i \le n}\{T_i\}$, for all patients. 
Then all survival times will fall into one of the $K$ disjoint intervals $I_k = [c_{k-1}, c_k)$, $k=1,\ldots, K$.   
In the interval $I_k$, $h_k= H_0(c_k) - H_0(c_{k-1})$ is the increment of the cumulative baseline hazard $H_0$. 
From the gamma process prior of $H_0$, the $h_k$'s have independent gamma distributions, i.e.
$
h_k \sim \mathcal{G}\left(\alpha_k - \alpha_{k-1}, a_0\right),
$
where $\alpha_k = a_0H^*(c_k)$. 
\cite{Ibrahim1999} showed numerically that this discrete approximation to an underlying gamma process prior is quite robust in many situations. 
Finally, we can obtain the grouped data likelihood
\begin{align}\label{formula:likelihood}
 L(\mathfrak D \mid \bm{\beta}, \bm h) \propto
 \prod_{k=1}^K \left[ \exp\left(-h_k \sum_{l\in \mathcal{R}_k \backslash \mathcal{D}_k} \exp\left(\mathbf X_l \bm{\beta} \right)\right)
 \prod_{\ell\in \mathcal{D}_k} ( 1 - \exp\left(-h_k \exp(\mathbf X_\ell \bm{\beta} )  )\right)   \right],
\end{align}
where $\bm h=(h_1,...,h_K)^\top$ and $\mathcal{D}_k$ is the failure set in the interval $I_k$, that is, the patients having the event during the interval. 

In order to investigate the effects of high-dimensional covariates ($p \gg n$), the stochastic search variable selection (SSVS) \citep{George1993} can identify a few important covariates and infer the effects of the identified covariates. 
The SSVS uses a prior for the regression coefficients $\beta$ called the ``spike-and-slab'' prior, which is a mixture of two normal distributions, one with a very large variance and one with a small variance, i.e.
\begin{equation}
\beta_j \mid \gamma_j \sim (1-\gamma_j) \mathcal{N}(0, \tau^2) + \gamma_j \mathcal{N}(0, c^2\tau^2), \ \ j = 1,...,p,
\end{equation}
where $\gamma_j$ acts as an inclusion indicator, so $\gamma_j = 0$ indicates that the $j$th covariate is not identified/included in the model, 
while $\gamma_j = 1$ indicates the exclusion of the $j$th variable in the model.
The scalar $\tau^2$ is set to be small so that $\beta_j$ is likely to be very close to $0$ when $\gamma_j = 0$, and $c^2\ (>1)$ is set to be large so that $\beta_j$ is likely to be non-zero when $\gamma_j = 1$.
A common choice for the prior of the variable selection indicators $\bm{\gamma} = (\gamma_1, \ldots, \gamma_p)^\top$ is the product of independent Bernoulli distributions \citep{Treppmann2017}, which leads to independent modeling of the covariates. 
In order to introduce dependence between the covariates for variable selection, we will use a Markov random field (MRF) prior instead, i.e.

\begin{equation}
f(\bm{\gamma} \mid \bm{G}) = \frac{\exp(a\boldsymbol{1}_{p}^\top \bm\gamma + b\bm{\gamma}^\top \bm G \bm{\gamma)}}{\displaystyle\sum_{\bm{\gamma}\in\{0,1\}^{p}}\bm{\gamma} + b\bm{\gamma}^\top \bm{G\gamma}} \propto \exp(a\boldsymbol{1}_{p}^\top \bm{\gamma} + b\bm{\gamma}^\top \bm G \bm{\gamma})
\end{equation}
where $\bm G$ is the adjacency matrix representation of a graph, and $a$ and $b$ are scalar hyperparameters. The overall inclusion probability is controlled by $a$ ($a<0$), while $b$ ($b>0$) controls how much the network information in the graph is weighted. 
Note that by using an empty graph in the MRF prior (or $b=0$), the model will be equivalent to using a product of independent Bernoulli distributions as the prior for the variable selection indicators $\bm{\gamma}$. 

\begin{figure}[h!tbp]

\newsavebox\tempbox
\sbox{\tempbox}{%

    \begin{tikzpicture}[node distance={15mm}, thick, main/.style = {draw, circle}] 
    \node[main] (1) {$X_1$}; 
    \node[main] (2) [above right of=1] {$X_2$}; 
    \node[main] (3) [below right of=1] {$X_3$}; 
    \node[main] (4) [above right of=3] {$X_4$}; 
    \node[main] (5) [above right of=4] {$X_5$}; 
    \node[main] (6) [below right of=4] {$X_6$}; 
    
    \draw (1) -- (2);
    \draw (1) -- (3);
    \draw (3) -- (4);
    \draw (4) -- (5);
    \draw (5) -- (6);
    \end{tikzpicture} 
}

\begin{subfigure}{.5\textwidth}
\centering
\usebox{\tempbox}
\caption*{}
\end{subfigure}%
\begin{subfigure}{.5\textwidth}
\centering
\vbox to\ht\tempbox{
    \vspace{-3mm}
    \begin{equation*}
    \bm G = 
    \begin{blockarray}{ccccccc}
     & \gamma_{1} & \gamma_{2} & \gamma_{3} & \gamma_{4} & \gamma_{5} & \gamma_{6}   \\
    \begin{block}{c(cccccc)}
      \gamma_{1} & 0  & 1  & 1  & 0  & 0 & 0    \\
     \gamma_{2}  & 1  & 0  & 0  & 0  & 0 & 0   \\
     \gamma_{3}  & 1  & 0  & 0  & 1  & 0 & 0   \\  
     \gamma_{4}  & 0  & 0  & 1  & 0  & 1 & 0    \\
     \gamma_{5}  & 0  & 0  & 0  & 1  & 0 & 1    \\
     \gamma_{6}  & 0  & 0  & 0  & 0  & 1 & 0    \\
    \end{block}
    \end{blockarray}
    \end{equation*}
}
\caption*{}
\end{subfigure}%
\caption{Illustration of a graph and its corresponding adjacency matrix.}
\label{fig:mrf_illustration}
\end{figure}
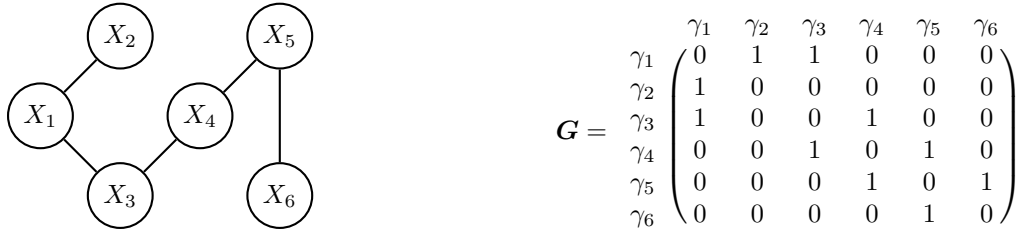

The MRF prior has been used to identify biological networks that can predict clinical outcomes of interest \citep{Stingo2011,Madjar2021}. 
For example, \cite{Madjar2021} used structure learning to estimate the graph in the MRF prior. 
However, structure learning requires a lot of data to be able to learn the graph well. 
In contrast, a fixed graph can be useful when substantial biological knowledge is already well-established, and can be especially helpful in the situation of low sample size and to save computational time. 
In this article, we will suggest a fixed graph for the prior based on a known biological network between the covariates in the penalized semiparametric Bayesian Cox model. 
This is similar to the use of an MRF prior in \cite{Stingo2011} and \cite{Zhao2021,Zhao2024jrssc}, but \cite{Stingo2011} used a parametric survival model and \cite{Zhao2021,Zhao2024jrssc} modeled continuous responses. 
Figure \ref{fig:mrf_illustration} shows an illustration of the graphical relationship between six variables and its corresponding adjacency matrix $\bm G$. 
Note that $\bm G$ is not necessary to be a proper (unweighted) adjacency matrix with all elements in $\{0,1\}$ in our proposed model; in fact, any element in $\bm G$ can be a non-negative real number to indicate the strength for joint inclusion of two variables \citep{Li2010, Zhao2024jrssc}. 

For the Bayesian inference, we slightly modified the algorithm from \cite{Madjar2021} by fixing the graph in the MRF prior (without graphical structure learning). 
The following MCMC Algorithm\ref{alg1} is to update the main parameters of interest, see Supplementary material S1 for more details.

\begin{algorithm}[H]
  \caption{MCMC algorithm}\label{alg1}
  \begin{algorithmic}[1]
   \State Set a graph $\bm G$ for the MRF prior and other hyperparameters. 
   \State Set initial values for $\bm{\beta}$, $\bm{h}, \bm{\gamma}$. 
   \State Set the number of iterations $L$.
   \For{\texttt{$l = 1,\cdots,L$ }}
      \State Sample variable selection indicators from $f(\gamma_j \mid \bm{\gamma}_{-j}, \beta_j, \bm{G})$ via Gibbs sampler.
      \State Sample regression coefficients from $f(\beta_j \mid \bm \beta_{-j}, \bm \gamma, \bm h, \mathfrak{D})$ via a random walk Metropolis-Hastings 
      \Statex\hspace{\algorithmicindent}algorithm with an adaptive jumping rule \citep{Lee2011}. 
      \State Sample increments of the cumulative baseline hazard from $f(h_k \mid \bm h_{-k}, \bm \beta, \bm\gamma, \mathfrak{D})$ via Gibbs sampler.
  \EndFor
  \end{algorithmic}
\end{algorithm}

We use the median probability model (MPM) to determine the finally selected variables and estimated values of coefficients \citep{Barbieri2004}. 
The MPM keeps only those covariates with marginal posterior inclusion probability larger than $0.5$, which has been shown to be the optimal model for prediction using a squared error loss function, under certain conditions, and which often yields good prediction performance even if the conditions are not met \citep{Barbieri2021}. 
The coefficients of the included covariates are estimated as the conditional posterior mean, and the coefficients of the excluded covariates are set to $0$, i.e.
\begin{equation}
\label{eq:mpm}
    \hat{\beta}_{j}^{\text{MPM}} = \begin{cases}
            \frac{\sum_{l=1}^L \beta_j^{(l)}}{\sum_{l=1}^L \gamma_j^{(l)}},\; & \text{if}\; \frac{\sum_{l=1}^L \gamma_j^{(l)}}{L} > 0.5, \\
            0, & \text{otherwise,}
    \end{cases}
\end{equation}
where $\beta_j^{(l)}$ and $\gamma_j^{(l)}$ are the estimates of the $l$th MCMC iteration. 
In this way, we select covariates that are the most frequently included in the MCMC simulation, and obtain a single model. 
An R package {\bf BayesSurvive} \citep{Zhao2024bayessurv} for implementing the proposed method is available on the Comprehensive R Archive Network at \url{https://CRAN.R-project.org/package=BayesSurvive}.

\subsection{Model assessment}

As we are interested in identifying stable covariates associated with survival, it is important to assess the stability of the variables selected in the model. 
The model size of a MPM can indicate how many stable covariates in the model are identified. 
If we know the truly relevant covariates, for example in a simulation study, we can use sensitivity (i.e. true positive rate), specificity (i.e. true negative rate) and accuracy to evaluate the variable selection performance of the MPM for the comparison of different approaches. 

We use the time-dependent Brier score (BS) and integrated Brier score (IBS) to evaluate the prediction performance of the models \citep{Brier1950,Graf1999}. 
The time-dependent BS measures the accuracy of the prediction at a specific time point by comparing the predicted survival probability $\hat{S}(t\mid \mathbf X_i)$ of all patients to their observed survival status $\mathbbm{1}_{\{T_i > t\}}$, 
$$
BS(t) = \frac{1}{n} \displaystyle\sum_{i=1}^n \hat{w}_i(t) \cdot \left( \mathbbm{1}_{\{T_i > t\}} - \hat{S}(t\mid \mathbf X)  \right)^2.
$$
The weights $\hat{w}_i(t)$ are to account for bias coming from the censored data, defined as 
$
\hat{w}_i(t) = \frac{ \mathbbm{1}_{\{T_i \le t\}} \delta_i }{ \hat{G}(T_i) } + 
\frac{ \mathbbm{1}_{\{T_i > t\}} }{ \hat{G}(t) },
$
where $\hat{G}$ is the Kaplan-Meier (KM) estimate for the censoring times.
The time-dependent BS can be summarized into a single value as the IBS over a time interval $[0,t^*]$, $t^* > 0$, 
$$
IBS(t^*) = \frac{1}{t^*} \int_0^{t^*} BS(t) \; \text dt,
$$
which measures the global prediction performance over a time interval rather than just at single time points.

\section{Simulations}
\label{sec:simulations}

To investigate the impact of the graph in the MRF prior on the variable selection as well as on survival prediction, we design simulation data sets with fully or partially known graph/network information between all covariates, which allows us to see the impact of having fully versus only partially known network information. 
We designed two simulation studies. 
\hyperref[sec:simstudy1]{Simulation study I} is to compare Bayesian Cox models without known network information, with known all correct network information, with known partially correct network information, and with partially incorrect network information. 
\hyperref[sec:simstudy2]{Simulation study II} is to perform additional sensitivity analysis by using different levels of known partially correct or incorrect network information. 

In \hyperref[sec:simstudy1]{Simulation study I}, we designed different scenarios of partially correct network information, and compared those against the two extreme scenarios, i.e. empty graph and true graph. 
Using the empty graph in the MRF prior means that covariates will be modeled independently, i.e. with independent Bernoulli priors for the inclusion indicators $\gamma_j$ $(j = 1,...,p)$. 
Using the true graph means that the entire precision matrix used to generate covariates is transformed to an adjacency matrix for the MRF prior. 
Since it is usually unrealistic to assume that we have the full graph information in any real application, we will investigate how sensitive the model is to having only partially correct or even wrong graph information (by adding false edges, and removing edges uniformly from the true graph and remove block edges) compared to the completely correct graph information. 
In \hyperref[sec:simstudy2]{Simulation study II}, we especially considered additional partially correct graph information which do not have marginal correlations between part of the truly correlated covariates (see adjacency matrices in Supplementary Figure \ref{fig:mrf_graphsII}), since in \hyperref[sec:simstudy1]{Simulation study I} those partially correct graphs kept marginal correlations between any pair of covariates from the true graph (Figure \ref{fig:mrf_graphs}).

We simulate a data set with $n=100$ individuals and $p=200$ covariates (i.e $n<p$). 
Assume that the first $20$ covariates are truly associated with the survival outcomes, so the sparsity is $10$\%. 
Within the $20$ truly relevant covariates, the first $15$ are correlated with covariance $0.5$ between each other, while all other variables are uncorrelated. 
This results in the following block covariance matrix and its corresponding precision matrix:
\[
\bm{\Sigma} = 
\begin{bmatrix}
    \bovermat{length 15}{1 & 0.5 & \cdots & 0.5} & 
    \bovermat{length 185}{0 & \cdots & 0} \\
    0.5 & \ddots & & \vdots & \vdots & & \vdots\\
    \vdots  &  & \ddots & 0.5 & \vdots & & \vdots\\
    0.5 & \cdots & 0.5 & 1 & 0 & \ldots & 0\\
    0 & \cdots& \cdots & 0 & \ddots & & \vdots \\
    \vdots & & &\vdots & &\ddots & 0\\
    0 & \cdots & \cdots & 0 & \cdots & 0 & 1\\
\end{bmatrix}, \ \ 
%
    \bm{\Omega} = 
\begin{bmatrix}
    \bovermat{length 15}{1.875 & -0.125 & \cdots & -0.125} & 
    \bovermat{length 185}{0 & \cdots & \hspace{0.5em}0\hspace{0.5em}} \\
    -0.125 & \ddots & & \vdots & \vdots & & \vdots\\
    \vdots  &  & \ddots & -0.125 & \vdots & & \vdots\\
    -0.125 & \cdots & -0.125 & 1.875 & 0 & \ldots & 0\\
    0 & \cdots& \cdots & 0 & \ddots & & \vdots \\
    \vdots & & &\vdots & &\ddots & 0\\
    0 & \cdots & \cdots & 0 & \cdots & 0 & 1.875\\
\end{bmatrix}.
\]
Based on the precision matrix $\bm\Omega$, we can construct a true graph where edges between covariates represent non-zero entries in the precision matrix. 
By specifying the true graph in the Bayesian Cox model with MRF prior, we will study its impact on the variable selection and prediction performance. Likewise, we can study the impact of perturbations, i.e. removed true edges or added false edges, by specifying perturbed graphs.
The covariates vector of each individual is simulated from the multivariate normal distribution, 
$
    \mathbf X_i \sim \mathcal{N}(0, \bm{\Sigma}), \ (i=1,\ldots,n).
$
Combining covariates of all individuals together by rows into a matrix, we get the design matrix $\mathbf X$. 
The coefficients of the first 20 covariates are simulated from the uniform distribution $[-1,1]$, while all other coefficients are set to $0$. 
The survival outcomes are simulated based on the linear predictor $\mathbf X\bm\beta$ under the assumptions of the Cox model with a Weibull-distribution cumulative baseline hazard function, as in \cite{Madjar2021}. 

It is not straightforward to obtain suitable credibility intervals for the median probability model. 
We assess the uncertainty associated with sampling variability by repeatedly generating data $\mathbf X$ $20$ times (i.e. $20$ training sets, all based on the same fixed model parameters $\bm{\gamma}$ and $\bm{\beta}$) to obtain a $95\%$ confidence interval of each MPM's coefficient based on 20 fitted MPMs from the 20 training sets. 
Similarly we can assess the stability of MPM's variable selection.
To assess the performance of survival prediction, we simulate one independent test set with the same specification of parameters. 
The test set will be used for calculating the (time-dependent) Brier score for all models fitted from the $20$ training sets. 

\subsection{Simulation study {I}}
\label{sec:simstudy1}

\begin{figure}[h!tbp]
\centering
\includegraphics[width=0.55\textwidth]{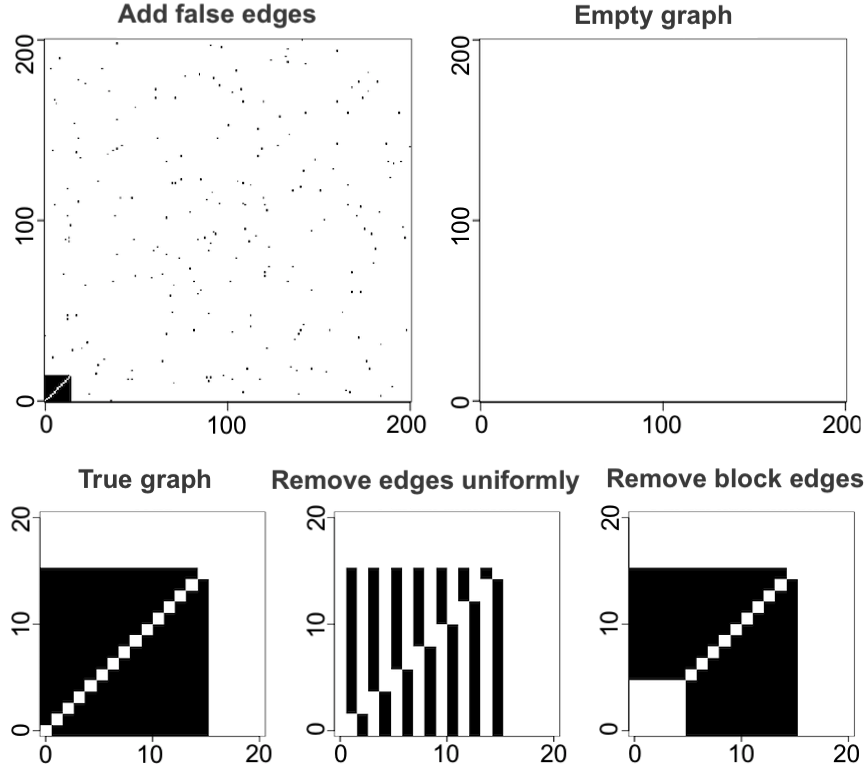}
\caption{Simulation scenario for Simulation study I: The adjacency matrices of the five different graphs used in the MRF prior in Simulation study I. 
Black entries indicate values of 1 and white entries indicate values of 0. 
Note that the bottom three panels only show the part of the adjacency matrix corresponding to the first $20$ truly relevant covariates, as all other entries are $0$.}
\label{fig:mrf_graphs}
\end{figure}

We compared the Bayesian Cox model with five different graphs in the MRF prior to investigate the impact of using different types of network information. 
The five graphs are defined as follows and Figure \ref{fig:mrf_graphs} shows the corresponding adjacency matrices of the five graphs. 

\begin{enumerate}[label={(\roman*)}]
    \item \textbf{Empty graph}: the empty graph contains no network information between the covariates, i.e. covariates are modeled independently. 
    The MRF prior with an empty graph is equivalent to using a product of independent Bernoulli distributions as the prior for the inclusion indicators $\bm{\gamma}$.
    
    \item \textbf{True graph}: this graph contains the covariance information that was used to generate the data set. 
    The precision matrix used for generating the covariates is turned into an adjacency matrix, meaning that every non-zero entry is set to $1$, and the diagonal is set to $0$. 

    \item \textbf{Remove edges uniformly}: $50\%$ of the edges are removed uniformly from the true graph. 
    The graph is constructed by vectorizing the adjacency matrix for the true graph, and then removing every second edge, and then transforming the vector back to a $p\times p$ matrix. 
    All the first $15$ covariates will keep their marginal correlations with each other, but they are no longer all direct neighbours.

    \item \textbf{Remove block edges}: $11\%$ of the edges are removed from the true graph. 
    This graph is constructed by setting a block of size $5$-by-$5$ where all entries were $1$ (apart from the diagonals) in the true graph to $0$. 
    Note that this also keeps marginal correlation between all the first $15$ covariates, but it severely thins out a part of the graph, meaning that the first $5$ variables will have fewer neighbours compared to those in the true graph.

    \item \textbf{Add false edges}: false edges are added randomly to the true graph. 
    We have added the same amount of false edges as the amount of true edges.

\end{enumerate}

\begin{figure}[!htpb]
    \centering
    \includegraphics[width=0.5\textwidth]{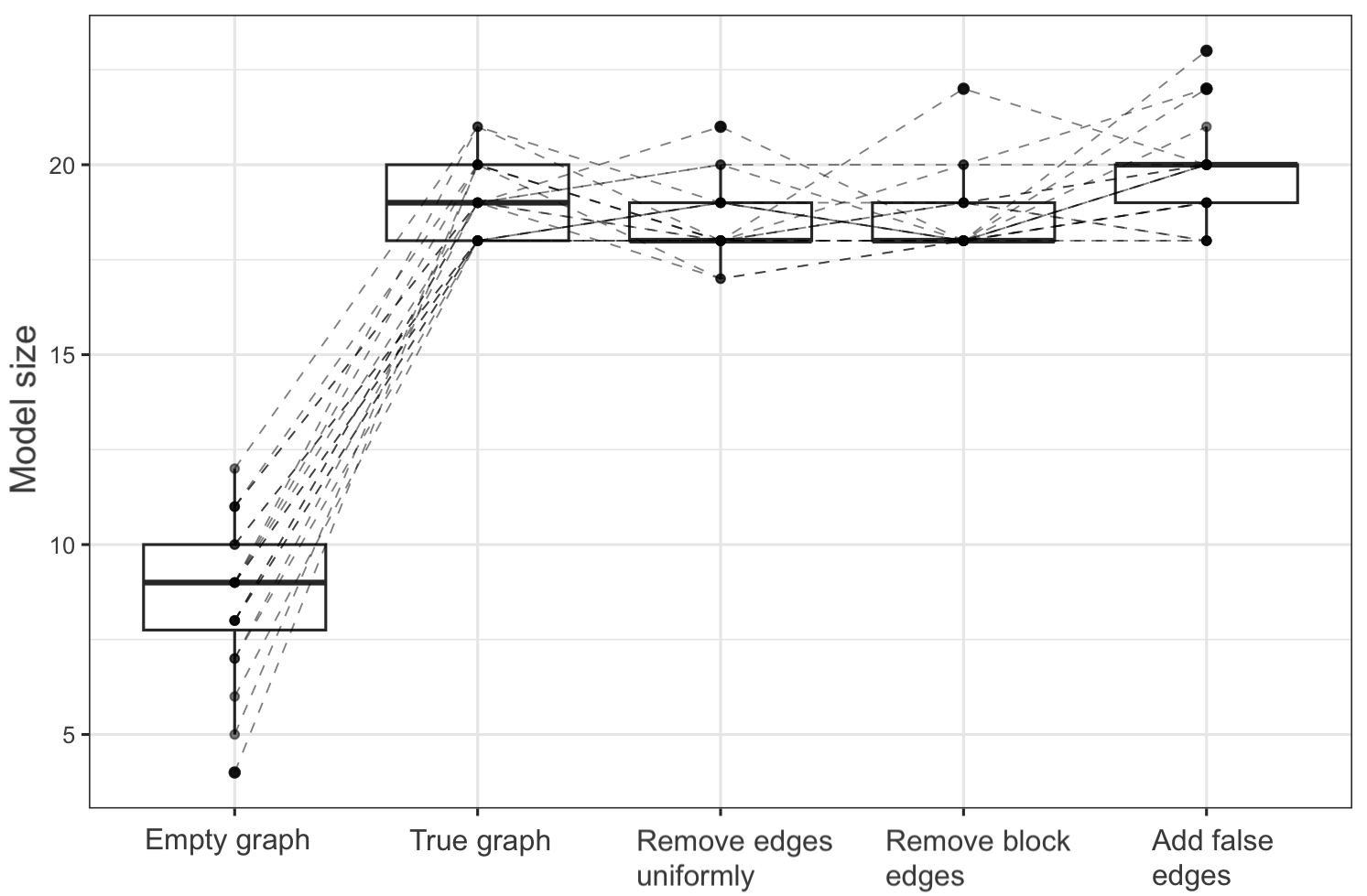}
    \caption{Results for Simulation study I: Model size of the median probability model for the five Bayesian Cox models for all 20 simulated data sets. The dotted lines are drawn between the same data sets.}
    \label{fig:model-size-boxplot}
\end{figure}

For each of the five different graphs in the MRF prior, we ran the Bayesian Cox model with $30,000$ MCMC iterations for each of the $20$ data sets. 
Since we used half of the iterations as the warmup period and a thinning parameter of $6$, there are $2500$ MCMC samples used for inference. 
The trace plots of the model size (Supplementary Figure \ref{fig:mcmc-model-size}) and the log-likelihood (Supplementary Figure \ref{fig:mcmc-loglik}) were used to investigate the mixing of the MCMC algorithm and convergence to the target distribution, which show the MCMC chains to be quite stable for these global summary statistics. 
Supplementary Figure \ref{fig:sim:eff_sample_size} shows the effective sample sizes for model size and log-likelihood. 
The effective sample sizes of the model with the empty graph in the MRF prior is quite small, as the modeling of the high-dimensional data is challenging in the presence of no useful prior knowledge and it resulted in unstable variable selection over the MCMC iterations, which is reflected in the highly fluctuating model size in Figure \ref{fig:model-size-boxplot} and very low sensitivity (true positive rate) of variable selection in Table \ref{tab:initial_metric_tables}. 

For the performance of variable selection, Figure \ref{fig:model-size-boxplot} shows that the model with the empty graph results in much smaller model sizes than the true model size of $20$ and Table \ref{tab:initial_metric_tables} shows its very low average sensitivity (0.362) for variable selection. 
All other models result in model sizes close to $20$ (Figure \ref{fig:model-size-boxplot}) and good inclusion/exclusion results for the covariates as reflected by sensitivity, specificity and accuracy (Table \ref{tab:initial_metric_tables}). 
In particular, the models where edges were removed, either by removing edges uniformly or removing block edges, performed very similarly to the model with the true graph, as did the model with the same amount of false edges added to the true graph. 

\begin{table}[!htpb]
    \centering
    \caption{Results for Simulation study I: Sensitivity, specificity and accuracy tables for the variable selection performance of the median probability model (MPM). The standard errors over the $20$ data sets are provided in parenthesis next to the mean value.}
    \label{tab:initial_metric_tables}
\begin{tabular}{llll}
\hline
\hline
  & Sensitivity & Specificity & Accuracy\\
\hline
Empty graph & 0.362 (0.084) & 1.000 (0.001) & 0.936 (0.009)\\

True graph & 0.900 (0.016) & 0.999 (0.003) & 0.989 (0.003)\\

Remove edges uniformly & 0.900 (0.016) & 0.999 (0.003) & 0.989 (0.003)\\

Remove block edges & 0.900 (0.016) & 0.999 (0.003) & 0.989 (0.003)\\

Add false edges & 0.900 (0.016) & 0.998 (0.004) & 0.988 (0.004)\\
\hline
\hline
\end{tabular}
\end{table}

\begin{figure}[!hp]
    \centering
    \includegraphics[width=1.\textwidth]{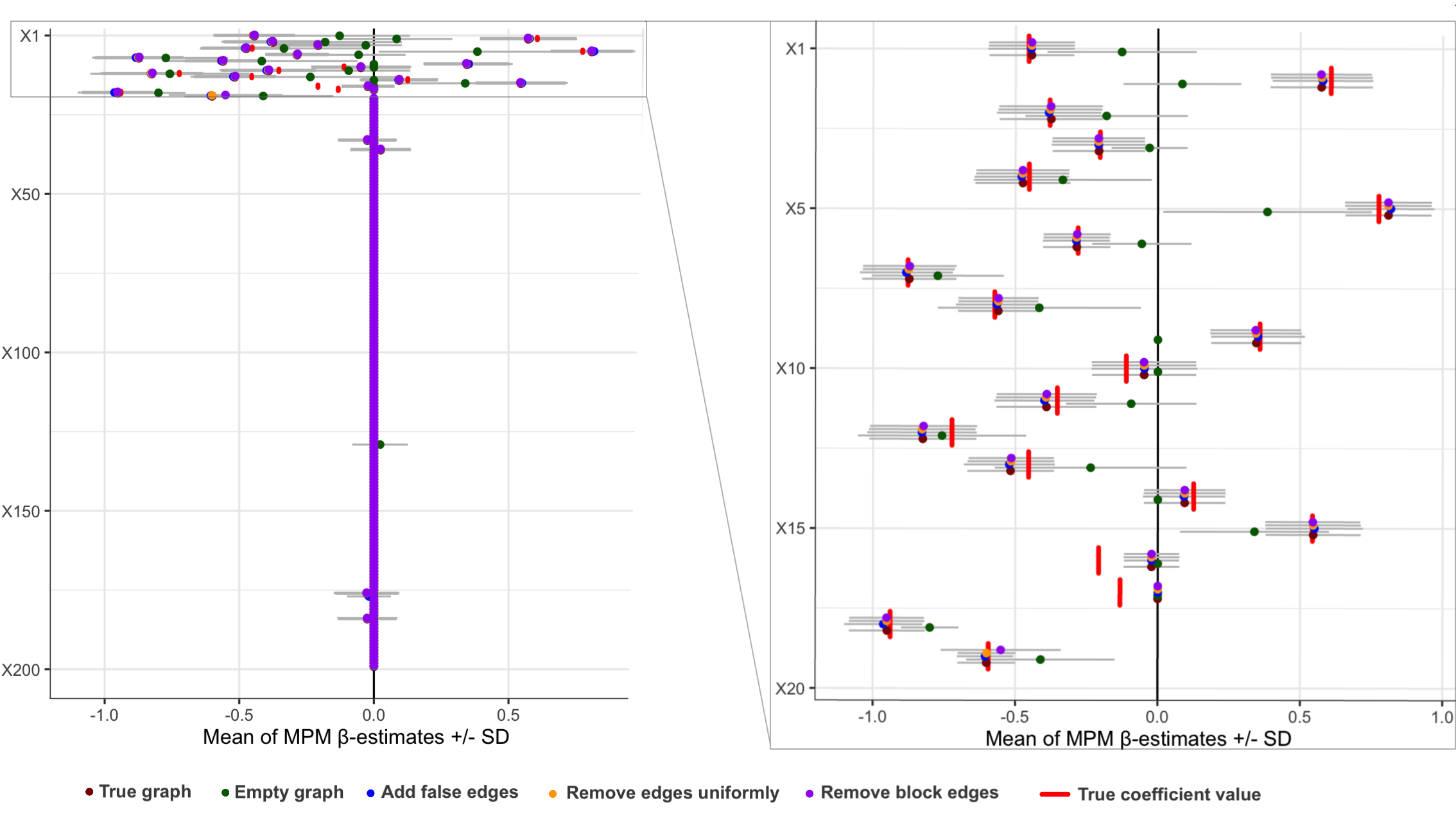}
    \caption{Results for Simulation study I: Estimated coefficients for the median probability model (MPM). The gray bar shows the averaged MPM $\beta$-estimate $\pm$ standard deviation (SD) over the $20$ simulated data sets. 
    The right panel shows the estimates of only the $20$ truly relevant covariates for clarity.}
    \label{fig:beta-mpm-full}
\end{figure}

To look at the median probability model (MPM) estimates of regression coefficients, Figure \ref{fig:beta-mpm-full} shows that the coefficients of most of the first $20$ truly relevant covariates are far from zero and that those of the $180$ truly irrelevant covariates are very close to zero, as expected. 
Of the five models, only for the model with the empty graph (green dots in Figure \ref{fig:beta-mpm-full}) some of the coefficient estimates are far away from the true coefficient values. 

\begin{figure}[!htbp]
    \centering
    \includegraphics[width=0.5\textwidth]{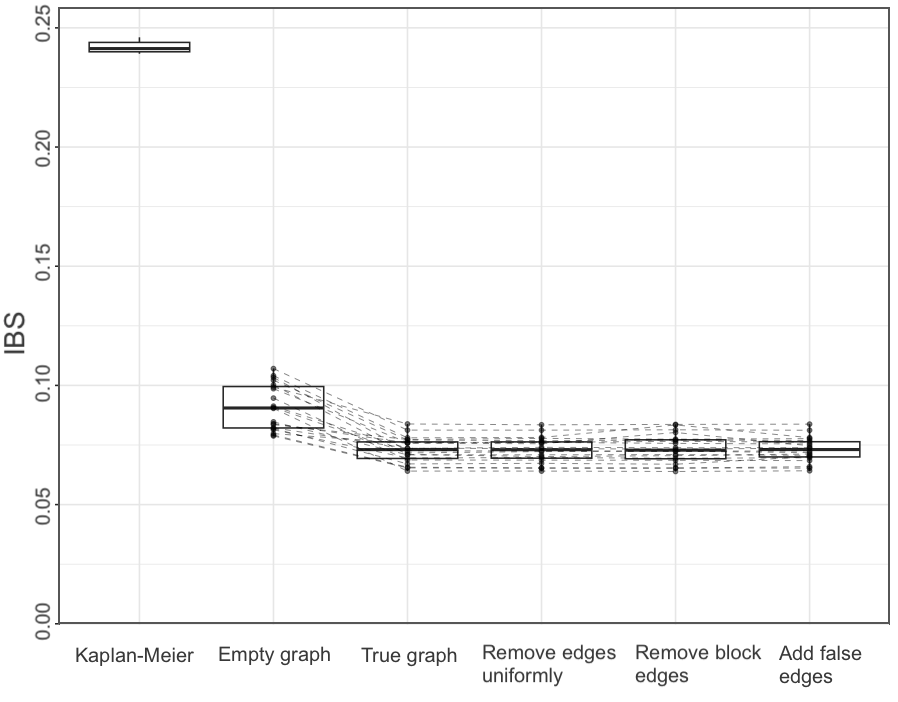}
    
    \caption{Results for Simulation study I: Integrated Brier scores (IBS) of the models fitted on the 20 training data sets evaluated for the independent test set based on the median probability models for the five different graphs. Dotted lines are connecting the results for the same training data sets.}
    \label{fig:ibs-boxplot}
\end{figure}

For the prediction of survival outcomes, Figure \ref{fig:ibs-boxplot} shows that while the model using the empty graph has smaller integrated Brier scores (IBS) than the Kaplan-Meier method, the IBS values are larger (indicating worse prediction performance) than for the models with the other four graphs. 
The prediction performances in terms of IBS are similar for all the models with the non-empty graphs, both the true graph and the partially true graphs. 
Supplementary Figure \ref{fig:time-dep-brier} also shows that the predicted time-dependent Brier scores for the model with the empty graph are consistently worse than for the models with the other four graphs at different evaluation times. 

\vspace{-2mm}
\subsection{Simulation study {II}}
\label{sec:simstudy2}

In the previous section, we observed that for the three models with different partially correct network information, the variable selection and survival prediction performances were all similar to the model using the true graph in the MRF prior. 
In this simulation study, we further look into each of the three notions of partially correct network information that were introduced in \hyperref[sec:simstudy1]{Simulation study I}, and compare different levels of information in these partially informative graphs to the fully non-informative (empty) and the fully informative (true) graphs. 

In addition to removing $50\%$ edges from the true graph uniformly by keeping every $k=2$nd edge and removing the rest, we here also choose $k=4, 6, 9$ to achieve $75\%$, $86\%$ and $94\%$ removal of true edges, respectively. 
And for the removal of block edges, in addition to removing a block of size $5\times 5$ of edges in \hyperref[sec:simstudy1]{Simulation study I}, here we also remove a block of size $10\times 10$ of edges from the adjacency matrix of the true graph. 
However, any two truly correlated variables in the above two scenarios still result in marginal correlation in the graph. 
Two new scenarios for removing block edges are therefore included here, where the marginal correlation is removed between some of the first 15 truly relevant covariates. 
For the graph with added false edges, in addition to adding the same (i.e. $100\%$) amount of false edges as the amount of true edges as in \hyperref[sec:simstudy1]{Simulation study I}, here we also include scenarios where $50\%$ or $200\%$ false edges are added, respectively. 
All scenarios are summarized as follows:

\begin{itemize}
    \item \textbf{Partial graph: remove edges uniformly}
   	 \begin{itemize}
            \item \textbf{50\% removal:} we keep every second edge from the true graph ($k=2$), which is the same as the graph simply named ``remove edges uniformly'' in \hyperref[sec:simstudy1]{Simulation study I}. 
            
            \item \textbf{75\% removal:} we keep every fourth edge from the true graph, so $k=4$.
        
            \item \textbf{86\% removal:} we keep every sixth edge from the true graph, so $k=6$, 
        
            \item \textbf{94\% removal:} we keep every ninth edge from the true graph, so $k=9$. 
	\end{itemize}

    \item \textbf{Partial graph: remove block edges}
        \begin{itemize}
            \item \textbf{5x5:} this graph is constructed by removing a block of size $5\times 5$ of edges from the adjacency matrix of the true graph (removing 11\% edges). This is the same as the graph simply named ``remove block edges'' in \hyperref[sec:simstudy1]{Simulation study I}. 
            
            \item \textbf{10x10:} this graph is constructed by removing a block of size $10\times 10$ of edges from the adjacency matrix of the true graph (removing 44\% edges).
        
            \item \textbf{5x5 plus:} this graph is constructed in an identical way to the previous graph ``remove block edges 5x5'', but we disconnect the $5$ covariates from all their neighbours (removing 56\% edges), see Supplementary Figure \ref{fig:mrf_graphsII}C. This is done by setting the rows and columns corresponding to these covariates to zero in the adjacency matrix.
            
            \item \textbf{10x10 plus:} this graph is constructed in an identical way to the previous graph ``remove block edges 10x10'', but we disconnect the $10$ covariates from all their neighbours (removing 89\% edges), see Supplementary Figure \ref{fig:mrf_graphsII}D.
        \end{itemize}
        
    \item \textbf{Noisy graph: add false edges}
        \begin{itemize}
            \item \textbf{50\%:} we randomly add false edges to the true graph. The number of false edges added is half of the number of edges already present in the true graph.
        
            \item \textbf{100\%:} we randomly add false edges to the true graph. The number of false edges added is equal to the number of edges already present in the true graph. This was the graph used as the noisy graph in \hyperref[sec:simstudy1]{Simulation study I}.
        
            \item \textbf{200\%:} we randomly add false edges to the true graph. The number of false edges added is twice the number of edges already present in the true graph. 
        \end{itemize}

\end{itemize}


\begin{figure}[!htbp]
    \centering
    \includegraphics[width = 0.7\textwidth]{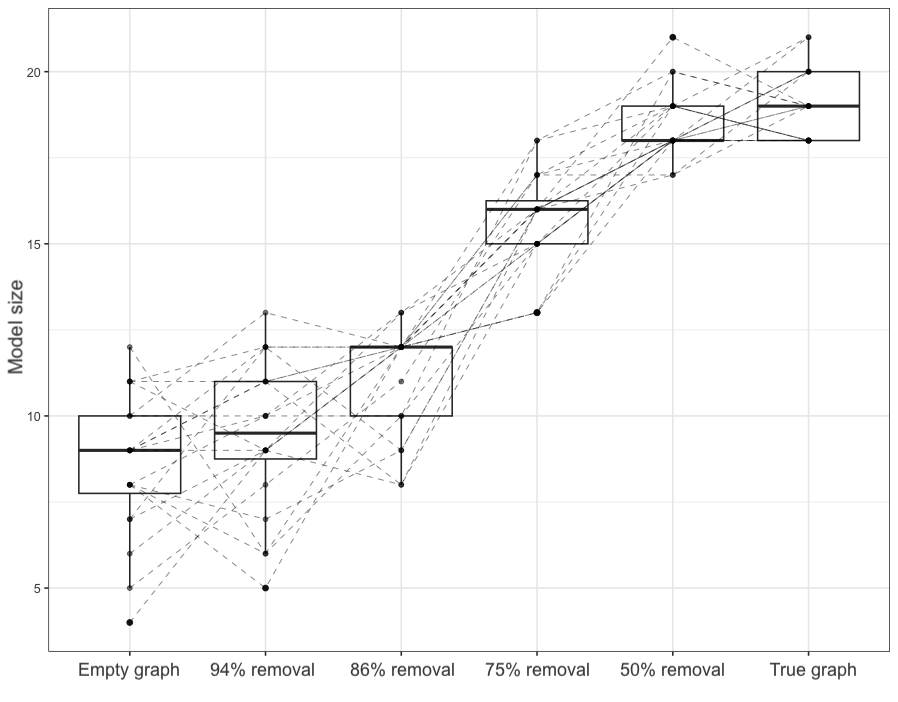}
    \caption{Results for Simulation study II: Model size of the Bayesian Cox models with MRF prior with different graphs. Each dot is the model size of the resulting median probability model for each of the 20 training sets, and the dotted lines connect the model sizes belonging to the same training set across different models}
    \label{fig:uniform_model_size}
\end{figure}

Figure \ref{fig:uniform_model_size} shows that there is certainly a trend to observe larger model sizes for the models where the graphs contain more information, i.e. are closer to the true graph, which is consistent with the observed sensitivities in Table \ref{tab:simulation2_metrics} (Partial graph: remove edges uniformly).  
We can observe this in the model sizes in Figure \ref{fig:uniform_model_size}, where we can also see that there is a larger jump from $86$\% removal to $75$\% removal, while at $50$\% removal the model size is quite similar to the model with the true graph in the MRF prior. 
Supplementary Figure \ref{fig:uniform_ibs} shows that the survival prediction IBS becomes worse when uniformly removing more and more correct edges in the graph information, and the uncertainty also becomes larger. 
For the prediction performance in terms of IBS, it seems that already at $75$\% removal the model performs almost indistinguishable from the model with the true graph in the MRF prior. 

\begin{table}[!htbp]
    \centering
    \caption{Results for simulation study II: Sensitivity, specificity and accuracy for the models with different graphs. The standard error across the $20$ data sets are provided in parenthesis next to the mean value.}
    \label{tab:simulation2_metrics}
    \begin{tabular}{llll}
\hline
\hline
  & Sensitivity & Specificity & Accuracy\\
\hline
True graph & 0.9 (0.016) & 0.999 (0.003) & 0.989 (0.003)\\

Empty graph & 0.362 (0.084) & 1 (0.001) & 0.936 (0.009)
\medskip\\

{\bf Partial graph: remove edges uniformly}\\
\hspace{2mm} Remove edges uniformly 50\% & 0.9 (0.016) & 0.999 (0.003) & 0.989 (0.003)\\

\hspace{2mm} Remove edges uniformly 75\% & 0.8 (0.032) & 0.999 (0.004) & 0.979 (0.005)\\

\hspace{2mm} Remove edges uniformly 86\% & 0.545 (0.078) & 1 (0.001) & 0.954 (0.008)\\

\hspace{2mm} Remove edges uniformly 94\% & 0.405 (0.086) & 1 (0.001) & 0.94 (0.009)
\medskip\\

{\bf Partial graph: remove block edges}\\

\hspace{2mm} Remove block edges 5x5 & 0.9 (0.016) & 0.999 (0.003) & 0.989 (0.003)\\

\hspace{2mm} Remove block edges 10x10 & 0.895 (0.022) & 0.999 (0.004) & 0.988 (0.005)\\

\hspace{2mm} Remove block edges 5x5 plus & 0.71 (0.053) & 1 (0.001) & 0.971 (0.006)\\

\hspace{2mm} Remove block edges 10x10 plus & 0.547 (0.068) & 0.999 (0.002) & 0.954 (0.007)
\medskip\\

{\bf Noisy graph: add false edges}\\

\hspace{2mm} Add false edges 50\% & 0.895 (0.022) & 0.998 (0.005) & 0.988 (0.005)\\

\hspace{2mm} Add false edges 100\% & 0.9 (0.016) & 0.998 (0.004) & 0.988 (0.004)\\

\hspace{2mm} Add false edges 200\% & 0.897 (0.02) & 0.998 (0.004) & 0.988 (0.004)\\

\hline
\hline
\end{tabular}
\end{table}

For the variable section by the models with removing block edges from the true graph to different extents, Supplementary Figure \ref{fig:non-uniform_model_size} shows a clear trend of increased model size with increasing correct knowledge in the graph. 
The two models with graphs where some variables are completely disconnected from all other variables in the graph (i.e. remove block edges 5x5 plus and 10x10 plus), have smaller model size compared to those two models that still keep marginal correlation between all the first $15$ correlated truly relevant covariates. 
Table \ref{tab:simulation2_metrics} shows larger sensitivity for variable selection for the models with removing block edges than the model with the empty graph, even for the two cases where marginal correlation is lost for some variables (i.e. remove block edges 5x5 plus and 10x10 plus). 
All the four models with partially correct graphs by removing block edges from the true graph result in better IBS than the model with the empty graph (Supplementary Figure \ref{fig:non-uniform_ibs}).

For the models with the noisy graph, i.e. where different numbers of false edges were added randomly, their variable selection performances (Table \ref{tab:simulation2_metrics} (noisy graph) and Supplementary Figure \ref{fig:noise_model_size}) and survival prediction performances (Supplementary Figure \ref{fig:noise_ibs}) are very similar to the model with the true graph. 


\section{Application}\label{sec:application}

\subsection{TCGA data}

We apply the proposed method on the survival and multi-omics data for breast cancer basal subtype patients from The Cancer Genome Atlas (TCGA). 
We filtered female cancer patients with times from diagnosis to death at least 30 days and 187 patients remained (censoring rate $85.6\%$). 
The tumor tissue sample of every patient was measured by RNA-seq (Illumina HISEQ 2000 RNA sequencing) for gene expression and by whole-exome sequencing (Illumina Genome Analyzer IIx) for somatic mutations. 
By following the pipeline of \cite{Zhao2024tutorial}, we performed a $\log2$-transformation of the normalized gene expression counts and pre-filtered the most variable genes that together explain $10$\% of the cumulative variance over all patients, which resulted in $268$ out of $60,660$ gene expression features. 
For the mutation data, we included mutation features with prevalence between $5\%$ and $95\%$, i.e. with our data set of $187$ patients a mutation feature is included if the mutation is present in at least $10$ patients and at most $177$ patients, which resulted in $60$ out of $16,662$ mutation features. 
We also included the genomic features that are present in the Kyoto Encyclopedia of Genes and Genomes (KEGG) \citep{kegg} breast cancer (Homo sapiens) pathway (hsa05224) for our modeling and data analysis\footnote{Note that we gave the mutation features present in the KEGG pathway graph a more lenient limit, but we still require some prevalence, since if too few individuals have the mutation it is unlikely to be detected as an important feature by the model. A $2$\% prevalence limit with a data set of $187$ individuals results in a mutation feature present in the KEGG pathway graph needing to be present in at least $4$ individuals for the feature to be included in the model.}.  
Two clinical features, age and treatment, were also included, so the final covariates contain $414$ gene expression features, $75$ mutation features and $2$ clinical features, which adds up to $491$ features in total. 

\subsection{Construction of the MRF prior graph}

The KEGG breast cancer pathways contain verified biological mechanisms of breast cancer, so we are interested to see whether our approach can confirm multi-omics features that are present in the KEGG pathways and can potentially identify novel multi-omics features that are not present in the KEGG pathways to draw new biological knowledge. 
The graph of the MRF prior in the Bayesian Cox model can be constructed based on the KEGG pathways related to breast cancer (Homo sapiens), see Supplementary Figure \ref{fig:kegg_brca}. 
The KEGG pathways are directed graphs between proteins, but here our model needs an undirected graph between gene features (gene expressions and/or gene mutations). See Supplementary Figure \ref{fig:graph_construction} for an illustration on how we converted directed graphs of proteins into the corresponding graphs of gene features. 
Since we only used data from TCGA breast cancer basal subtype patients but the construction of the MRF prior graph is based on the KEGG pathways related to all subtypes of breast cancer, we can verify the robustness of our model toward noisy graph information.

We used the R package {\bf KEGGgraph} \citep{KEGGgraph} to parse the KEGG XML diagram into a graph object. 
Note that the nodes in the KEGG diagram are proteins and not genes, so we will also need to modify the nodes from {\bf KEGGgraph} to be gene features instead (one protein node can correspond to multiple genes, see Supplementary Figure \ref{fig:graph_construction}). 
We constructed the graph in such a way that each feature in the data matrix was mapped by Entrez Gene ID to the KEGG graph. 
If the corresponding protein of a feature is present in the KEGG graph, all its neighbours in the graph are checked. 
If the neighbour has a corresponding feature present in the data matrix, an undirected edge is added. 
We decided not to connect the different genes corresponding to the same protein together. 
This could lead to large fully connected blocks, which would encourage joint selection of large numbers of features, which is not warranted here given our limitations in sample size. 
Still, our approach leads to a quite informative graph with $635$ edges in total; the composition of the nodes and edges in the graph is shown in Supplementary Table \ref{tab:graph_composition}. 
Note that if a gene mutation feature and a gene expression feature correspond to the same gene, they are connected in the graph of the MRF prior, which follows the procedure by \cite{Zhao2024jrssc} for multi-omics integration. 

\subsection{Model setup and MCMC diagnostics}

We used a $90\%/10\%$ data split for the training and test sets. 
Because of the small number of events in the data, we split the censored and the non-censored observations into training and test sets separately, ensuring a fixed censoring proportion in each split. 
We repeated the random data split $20$ times to quantify the uncertainty of model performance, in particular measuring uncertainty of estimated MPM coefficients and the stability of variable selection by the Bayesian Cox models with the empty graph (denoted as BayesCox-Ber) and the KEGG graph (denoted as BayesCox-MRF).  
Note that we did not compare with the Bayesian Cox model with graphical learning proposed by \cite{Madjar2021}, since the computational cost of the model with graphical learning would be too high in our case, which contains $491$ covariates compared to only $20$ covariates in \cite{Madjar2021}. 

We ran $100,000$ MCMC iterations with half of the iterations as the warmup period. 
Due to the phase-transition phenomenon of the hyperparameters in the MRF prior, we tuned them according to the recommendation in \cite{Zhao2024jrssc}. 
In Supplementary Figure \ref{fig:mcmc_empty_real} and \ref{fig:mcmc_kegg}, we can see that the MCMC diagnostics plots for the model size show some autocorrelation, but the log-likelihoods look reasonably stable. 
Supplementary Figure \ref{fig:effective_sample_size}B shows acceptable effective sample sizes of the log-likelihoods, which supports the validity of the posterior inference of parameters of interest in the Bayesian Cox models. 

\subsection{Variable selection and prediction performance}

Our main goal is to achieve stable variable selection, with the ultimate aim to identify potential novel molecular biomarker candidates for personalized treatment strategies. 
Out of the $168$ genomic features that had at least one neighbour in the graph, $94$ of them were selected in at least $20$\% of the different data splits (referred to as stable variables). 
Table \ref{tab:in_out_of_graph} shows that the BayesCox-MRF model identified $94$ out of $180$ variables which are represented in the breast cancer KEGG pathways, while the BayesCox-Ber model did not find any of these KEGG breast cancer pathway variables as stable variables at all. 
The KEGG breast cancer pathways are well understood and have been verified extensively by biologists, which seems to translate to them being very informative with respect to survival. 
In addition, the likelihood of the model (\ref{formula:likelihood}) might only provide limited evidence here, as it is a challenging high-dimensional data set with a high censoring proportion and small sample size. 

\begin{table}[!ht]
\centering
\caption{TCGA data application: Number of variables selected stably, classified by whether or not they were present in the KEGG pathways related to breast cancer or in the KEGG basal subgraph.}
\begin{tabular}{lccc}
\hline
\hline
           				& KEGG & KEGG-basal & Others \\ \hline
Input data    			& 180        & 137    & 213     
\medskip\\
BayesCox-Ber 			& 94      	& 92    & 4  \\ 
BayesCox-MRF	        & 0      	& 0     & 1  \\ 
\hline
\hline
\end{tabular}

\label{tab:in_out_of_graph}
\end{table}

\begin{landscape}

\begin{figure}[!htbp]
\vspace{-15mm}
    \centering
    \makebox[\textwidth][c]{
    \includegraphics[width=1.\linewidth]{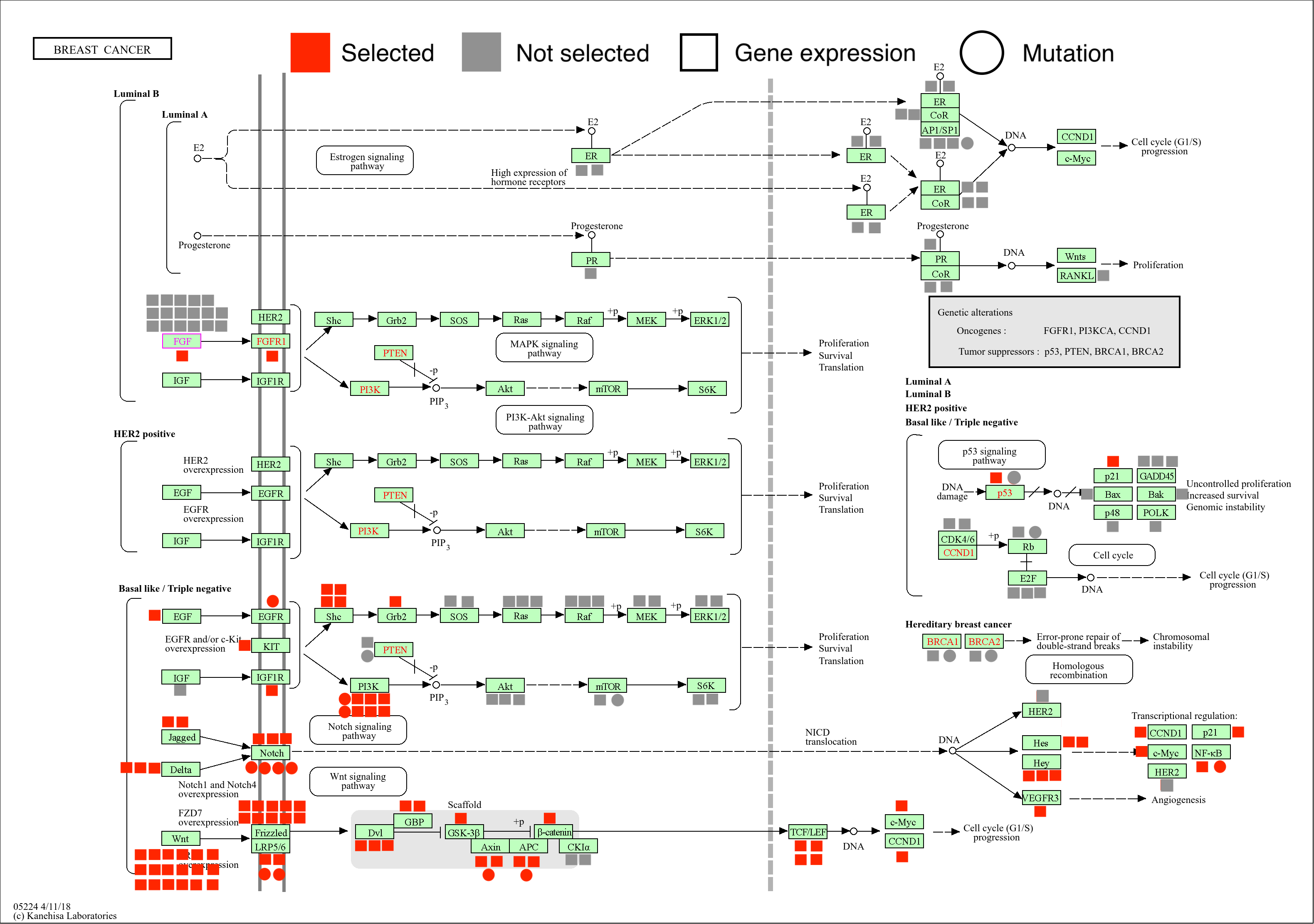} }
    \caption{TCGA data application: Visualization of the variable selection using the MPM with an informative MRF prior in the KEGG pathways related to breast cancer \cite{kegg}. A square or circle filled with red (gray) color represents a gene that was selected (not selected) corresponding to the protein/green-colored rectangle in the original KEGG pathways. 
    Note that there is some overlap in the pathways, with proteins featuring in multiple pathways for different subtypes. In the cases where a covariate corresponds to a protein feature in both the basal part and the non-basal part of the pathways, we only marked its location in the basal part. For example, the protein Grb2 features both in the HER2 positive subtype and the basal subtype, but we only marked it under the basal subtype part in the diagram.} 
    \label{fig:kegg_in_graph}
\end{figure}
\end{landscape}

Since we used the entire KEGG breast cancer pathways to construct the graph for the MRF prior, including the parts pertaining to breast cancer subtypes other than basal, it is relevant to assess which parts of the graph are represented by the selected variables. 
From Table \ref{tab:in_out_of_graph} and Figure \ref{fig:kegg_in_graph}, we see that the Bayesian Cox model with the KEGG graph was more likely to select a feature (i.e. gene expression or mutation feature) that corresponds to a protein in the basal part of the KEGG pathways, compared to a protein that did not appear in the basal part but rather belonged to a different breast cancer subtype. 
In particular, Figure \ref{fig:kegg_in_graph} shows that the genes that are up-stream in the basal-like pathways were all selected, while the genes down-stream in the pathways were selected less. 
This is in agreement with the assumption that the up-stream genes are most important, as they affect the whole pathway. 
Thus, we confirm that our model quite successfully filtered out the ``noise'', i.e. the biological connections belonging to other breast cancer subtypes, which are thought to be irrelevant for basal-like breast cancer. 

There are four novel genes (i.e. FBN3, CSAG3, SPRR2E, MAGEC2) selected from the gene expression features that did not appear in the KEGG pathways. 
Although we could not find any literature references linking these four genes to basal-like breast cancer specifically, there is some literature that indicates their importance for breast cancer prognosis in general. 
For example, \cite{Liu2020} found that FBN3 is related to prognosis of patients with breast cancer with bone metastasis. 
\cite{Yang2014} found that MAGEC2 positive expression has a worse prognosis and a shorter breast cancer metastasis-free survival. 
\cite{Qadir2018} found that cancer exosomes had significantly lower transactivation on SPRR2E expression when compared to normal exosomes. 
\cite{Shukla2018} found that both genes CSAG3 and MAGEC2 encoded cancer-germline antigens that are restrictively expressed in testis and placenta during normal development but re-expressed in many tumor types.

To assess the model performance of survival prediction, Supplementary Figure \ref{fig:ibs_real} shows that the integrative Brier scores (IBS) of both the BayesCox-Ber and BayesCox-MRF models are significantly smaller than of the reference frequentist Cox model with only clinical variables (i.e. age and treatment), which indicates that the multi-omics features do provide better prognostic power compared to simply using the two clinical variables alone. 
Even though our main goal in this application was improved variable selection and not prediction performance, it is still desired that the Bayesian Cox model with the KEGG graph MRF prior does not have worse prediction performance compared to the competing models. 
We noted that while the two Bayesian Cox models had better IBS than the Cox model with only the clinical variables, none of the Bayesian Cox models selected the clinical variables. 
We can see in Supplementary Figure \ref{fig:tcga_basal} that the treatment covariate has a large impact on survival, with a statistically significant log-rank test result. 
The reason why the clinical variables were not selected might be that they are correlated with other selected omics features and their prognostic value is therefore (partially) lost when including the omics features in the model. 
Therefore, it could be interesting to see how results might change if the Bayesian Cox model would allow the mandatory inclusion of important established (clinical) variables \cite{Zucknick2015}.  

\section{Discussion}
\label{sec:discussion}

We have presented a Bayesian Cox model for high-dimensional data, which allows the inclusion of known biological network information to guide the variable selection via Markov random field graph-structured selection priors (BayesCox-MRF). 
A main contribution of this manuscript is the systematic investigation of the impact of the MRF prior on model performance, both in terms of variable selection and prediction accuracy. 
We have investigated how to incorporate known biological network information into the MRF prior, and have performed extensive simulation studies and analyzed the TCGA basal-like breast cancer data set to compare the BayesCox-MRF model performance against models with an independent Bernoulli selection prior for all covariates (which corresponds to using an empty graph in the MRF prior; BayesCox-Ber). 
The main advantage of using known biological network information in the model is that, when biological research advances, the models could improve just from the improved network information provided, without any additional training data or improved methodology. 
The Bayesian Cox model could aid in identifying potentially novel biomarkers that can then be further studied by biologists. 

The simulation studies and the application to the TCGA breast cancer data showed notable benefits of using an MRF prior with known network information. 
In the simulation studies, the network information was crucial to be able to identify most of the truly relevant covariates, while the Bayesian Cox model with the empty graph had much lower sensitivity than even the model with partially correct graphs (Tables \ref{tab:initial_metric_tables} and \ref{tab:simulation2_metrics}). 
In the basal-like breast cancer data analysis, it was interesting to see that most of the parts of the KEGG breast cancer pathways that are directly associated with basal-like breast cancer were selected by the proposed Bayesian Cox model (Figure \ref{fig:kegg_in_graph} and Table \ref{tab:in_out_of_graph}). 
The graph constructed from the KEGG breast cancer pathways also contained pathways related to other subtypes, which are not thought to be relevant to basal-like breast cancer and were quite effectively filtered out by the model. 
Correspondingly, it was also indicated by the simulation studies that random noise (i.e. false edges in the network graph) did not affect variable selection or survival prediction much. 
That is, the Bayesian Cox model with an informative MRF prior is quite robust toward noise, i.e. added false edges in the graph.

There are several other possibilities for further work that could improve the proposed model and its implementation. 
Computational time will not only increase with the number of variables, but also with the number of partitions of the time axis, which is needed for generating the gamma process for the baseline hazard at every MCMC iteration. 
While it is not necessary to introduce a new partition at each event time point, a certain granularity is needed for a sufficiently accurate baseline hazard estimate, and therefore the number of partitions will tend to be associated with the number of events. 
Thus, as the number of distinct event times increases, this slows down the MCMC computations. 
Investigation into an efficient approach for speeding up the gamma process simulation, e.g. by approximations, could potentially reduce the computational costs for data sets with a large number of events. 
There are several other approaches that could be tried to increase computational efficiency. 
Variational inference is an approximation method to efficiently compute complex posterior probability distributions, in \cite{Komodromos2022} specifically for high-dimensional Bayesian proportional hazards models with spike-and-slab priors for variable selection. 
For high-dimensional covariates, the spike-and-slab priors usually lead to substantial computational challenges due to the combinatorial complexity of updating the discrete variable selection indicators \citep{Bhadra2019}, and the MRF prior can lead to phase transition problems. 
As an alternative, \cite{Denis2025} proposed the horseshoe Gaussian Markov random field prior (HS-GMRF), albeit so far only for linear models with continuous outcomes. 
Their prior setup is different from our approach in two ways. First, instead of discrete variable selection via a spike-and-slab prior, it uses a horseshoe prior which does not perform discrete variable selection, but instead induces continuous shrinkage of the regression coefficients. 
Consequently, \cite{Denis2025} then introduce a GMRF prior for continuous variables, i.e. for the regression coefficients, to allow local shrinkage parameters to capture the dependence between connected covariates, and which will encourage a similar amount of shrinkage for their regression coefficients. 
A standard horseshoe prior is used for non-connected variables. 
Thus, while this approach has the advantage of being scalable to high-dimensional covariate sets, it also encourages smoothness across the regression coefficients of connected covariates instead of simply encouraging joint selection. 
However, in applications in precision medicine, where the MRF graphs will often be built based on molecular pathways, the assumption that the regression coefficients of covariates within the same pathway should be similar will be too strong. 
 
Finally, since clinical prognostic factors are often important for the prediction of clinical outcomes and well established, one should demonstrate that the inclusion of genomic variables has added prognostic value over the established clinical prognostic factors \citep{DeBin2014}. Therefore, another useful extension of the proposed model would be to introduce so-called mandatory variables in the model, which are always included and not part of the variable selection \citep{Zucknick2015}. 


\section*{Data availability}

The TCGA breast cancer data were retrieved through the R/Bioconductor package TCGAbiolinks version 2.29.6 in February 2024. 
All code used for the simulation studies and the real data analysis can be found in the GitHub repository \url{https://github.com/tobiasoh/master_thesis}.


\bibliographystyle{apalike}
\bibliography{refs}

\clearpage

\setcounter{page}{1}
\pagenumbering{roman}

\section*{\centering Supplementary materials for \\``Bayesian Cox model with graph-structured variable selection priors for multi-omics biomarker identification''}


\affil[1]{Department of Mathematics,
University of Oslo, Oslo, Norway}

\affil[2]{Oslo Centre for Biostatistics and
Epidemiology, Department of Biostatistics,
University of Oslo, Oslo, Norway}
\renewcommand\thefigure{S\arabic{figure}}    
\setcounter{figure}{0}  
\renewcommand\thetable{S\arabic{table}}    
\setcounter{table}{0}

{\bf S1. Posterior inference}

The following list is an overview of the MCMC algorithm for updating the main parameters of interest. 
More details can be found in \cite{Madjar2021} but fixing the graph $\bm G$ in the MRF prior for variable selection indicator variables.

\begin{itemize}
    \item \textbf{Step 1:} Set the graph $\bm G$ of the MRF prior and other hyperparameters.
    
    \item \textbf{Step 2:} Set initial values for $\bm{\beta}$, $\bm{h}, \bm{\gamma}$. 
    
    \item \textbf{Step 3:} Update the inclusion indicators iteratively using the Gibbs sampler, through the conditional distribution 
    \begin{align*}
       f(\gamma_j \mid \bm{\gamma}_{-j}, \beta_j, \bm{G}) 
       &= \frac{f(\gamma_j, \beta_j\mid \bm\gamma_{-j}, \bm G)}{\sum_{l=1}^p f(\gamma_l, \beta_l \mid \bm\gamma_{-l} ,\bm G)} \nonumber \\
       &= \frac{f(\gamma_j \mid \bm\gamma_{-i}, \bm G) f(\beta_j \mid \bm \gamma, \bm G)}{\sum_{l=1}^p f(\gamma_l \mid \bm\gamma_{-l}, \bm G) f(\beta_l \mid \bm \gamma, \bm G)} \nonumber \\
        &= \frac{p(\bm \gamma \mid \bm G) f(\beta_j \mid  \gamma_j)}{\sum_{l=1}^p f(\bm \gamma \mid \bm G) f(\beta_l \mid  \gamma_l)},
    \end{align*}
    where $\bm\gamma_{-j}$ denotes the vector of $\bm\gamma$ excluding the $j$th element. Then, $\gamma_j = 1$ is accepted with probability 
    $$
        p(\gamma_j = 1 \mid \bm \gamma_{-j}, \bm G, \beta_j) = \frac{w_a}{w_a + w_b},
    $$
    where 
    \begin{align*}
        w_a &= \exp(a \boldsymbol{1}_p^\top \bm\gamma_ + b \bm\gamma^T \bm G \bm \gamma)\mid_{\gamma_j = 1} \cdot \mathcal{N}(\beta_j \mid 0, c^2 \tau^2), \nonumber \\
        w_b &= \exp(a \boldsymbol{1}_p^\top \bm\gamma_ + b \bm\gamma^T \bm G \bm \gamma)\mid_{\gamma_j = 0} \cdot \mathcal{N}(\beta_j \mid 0, \tau^2).
    \end{align*}
\end{itemize}

\begin{itemize}
    
    \item \textbf{Step 4:} Update $\beta_j$ for $j=1,\ldots, p$ from $f(\beta_j \mid \bm \beta_{-j}, \bm \gamma, \bm h, \mathfrak{D})$ using a random walk Metropolis-Hastings algorithm with an adaptive jumping rule \citep{Lee2011}. The conditional posterior distribution is 
    \begin{align*}
        &f(\beta_j \mid \bm\beta_{-j}, \bm \gamma, \bm{h}, \mathfrak{D})  \\
        \propto\ 
        &L(\mathfrak{D}\mid \bm\beta, \bm h) \cdot f(\bm\beta \mid \bm \gamma)  \\
        \propto\ &\prod_{k=1}^K \left[ \exp\left(-h_k \sum_{l\in \mathcal{R}_k \backslash \mathcal{D}_k} \exp\left(\mathbf X_l \bm{\beta} \right)\right)
 \prod_{\ell\in \mathcal{D}_k} ( 1 - \exp\left(-h_k \exp(\mathbf X_\ell \bm{\beta} )  )\right)   \right]  
        \cdot \exp(-\frac{1}{2} \bm\beta^\top \Sigma^{-1}\bm\beta),
    \end{align*}
	where $\bm\beta_{-j}$ denotes the vector of $\bm\beta$ excluding the $j$th element, and $\Sigma = \text{diag}\{\sigma_{\beta_1}^2,...,\sigma_{\beta_p}^2\}$ with $\sigma_{\beta_j}^2 = (1-\gamma_j)\tau^2 + \gamma_j c^2\tau^2$, $j=1,...,p$.

    \item \textbf{Step 5:} Update $h_k$ using the conditional distribution $f(h_k \mid \bm h_{-k}, \bm \beta, \bm\gamma, \mathfrak{D})$, which can be approximated by the gamma distribution 
    $$
        h_k \mid \bm h_{-k}, \bm\beta, \bm\gamma, \mathfrak{D}\;\;\; \overset{approx}{\sim}  
        \mathcal{G}\left(a_0(H^*(c_k) - H^*(c_{k-1})) + d_k,\;\; a_0 + \sum_{l\in \mathcal R_k \backslash \mathcal D_k} \exp(\mathbf  X_l\bm\beta)\right),
    $$
    where $\bm h_{-k}$ denotes the vector of $\bm h$ excluding the $k$th element, and $d_k$ is the number of individuals in the failure set $\mathcal D_k$.

\end{itemize}

In {\bf Step 4} the Metropolis-Hastings algorithm for the regression coefficients, the update in $l$th iteration is determined as follows:
\begin{enumerate}
    \item Sample a proposal $\beta_j^*$ from the adaptive proposal distribution $q(\beta_j^* \mid \beta_j^{(l-1)} = \mathcal{N}(\beta_j^*\mid u_{\beta_j}^{(l-1)}, v_{\beta_j}^{(l-1)})$, where $u_{\beta_j}^{(l-1)}$ and $v_{\beta_j}^{(l-1)}$ are the first and second derivatives of $\log f(\beta_j \mid \bm\beta_{-j}, \bm \gamma, \bm{h}, \mathfrak{D})$ with respect to $\beta_j$ evaluated with $\beta_j^{(l-1)}$.

    \item Calculate 
    \begin{equation}
        r = \frac{f(\beta_j^* \mid \bm\beta_{-j}^{(l-1)}, \bm\gamma^{(l-1)}, \bm h^{(l-1)}, \mathfrak{D})/q(\beta_j^* \mid \beta_j^{(l-1)})}{f(\beta_j^{(l-1)} \mid \bm\beta_{-j}^{(l-1)}, \bm\gamma^{(l-1)}, \bm h^{(l-1)}, \mathfrak{D})/q(\beta_j^{(l-1)} \mid \beta_j^*)}.
    \end{equation}

    \item Generate a random number $u \sim \text{Uniform}[0,1]$, and accept the proposal $\beta_j^*$ if $\min(1, r) > u$.
    
\end{enumerate}

\clearpage
{\bf S2. Additional simulation results}

\begin{figure}[h!tbp]
    \centering
    \textbf{Model size plots for MCMC diagnostics}\par\medskip    

\begin{minipage}{0.5\textwidth}
    \includegraphics[width=0.9\textwidth]{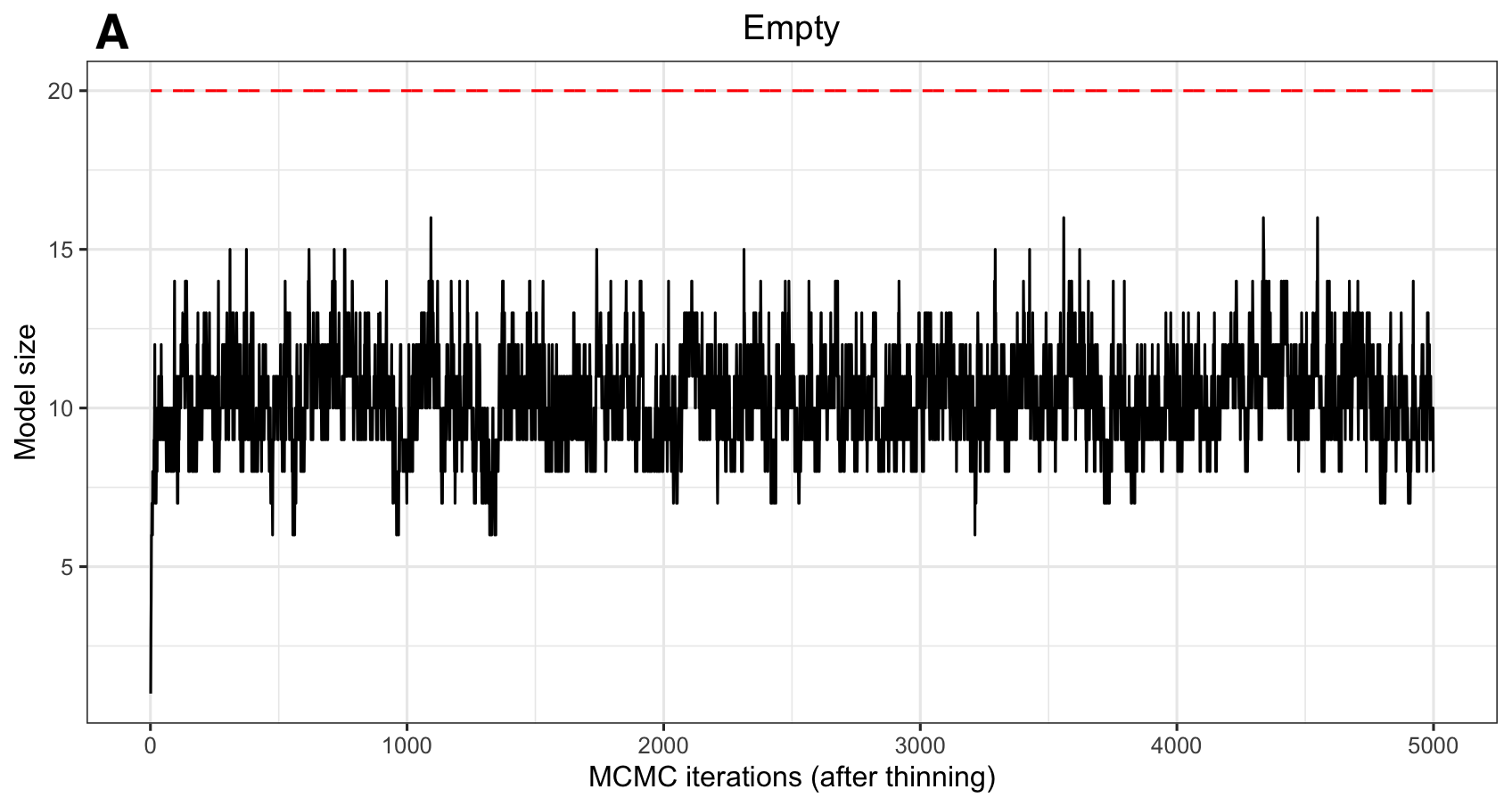}
\end{minipage}%
\hfill
\begin{minipage}{0.5\textwidth}
    \includegraphics[width=0.9\textwidth]{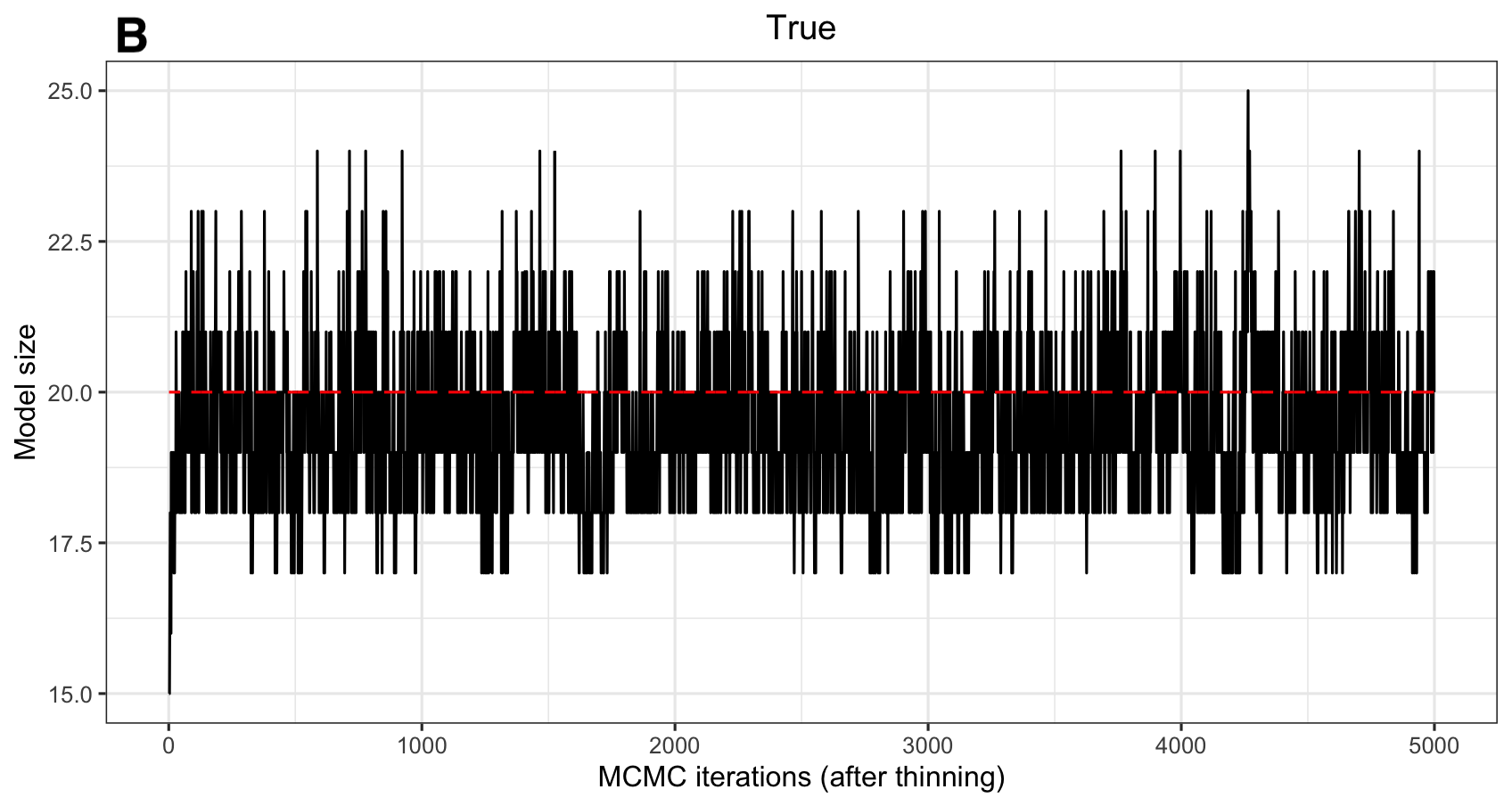}   
\end{minipage}

\begin{minipage}{0.5\textwidth}
    \includegraphics[width=0.9\textwidth]{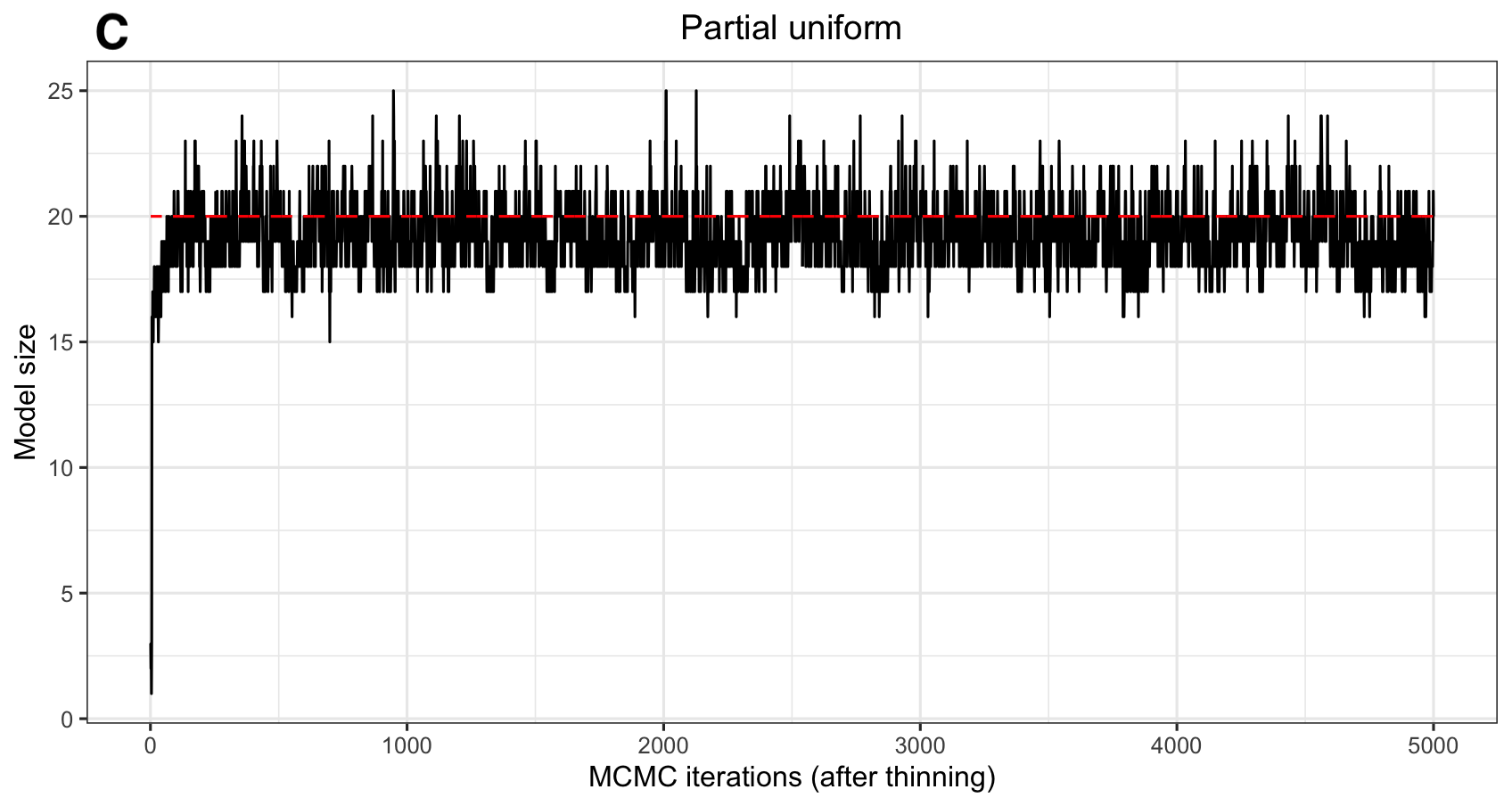}     
\end{minipage}%
\hfill
\begin{minipage}{0.5\textwidth}
    \includegraphics[width=0.9\textwidth]{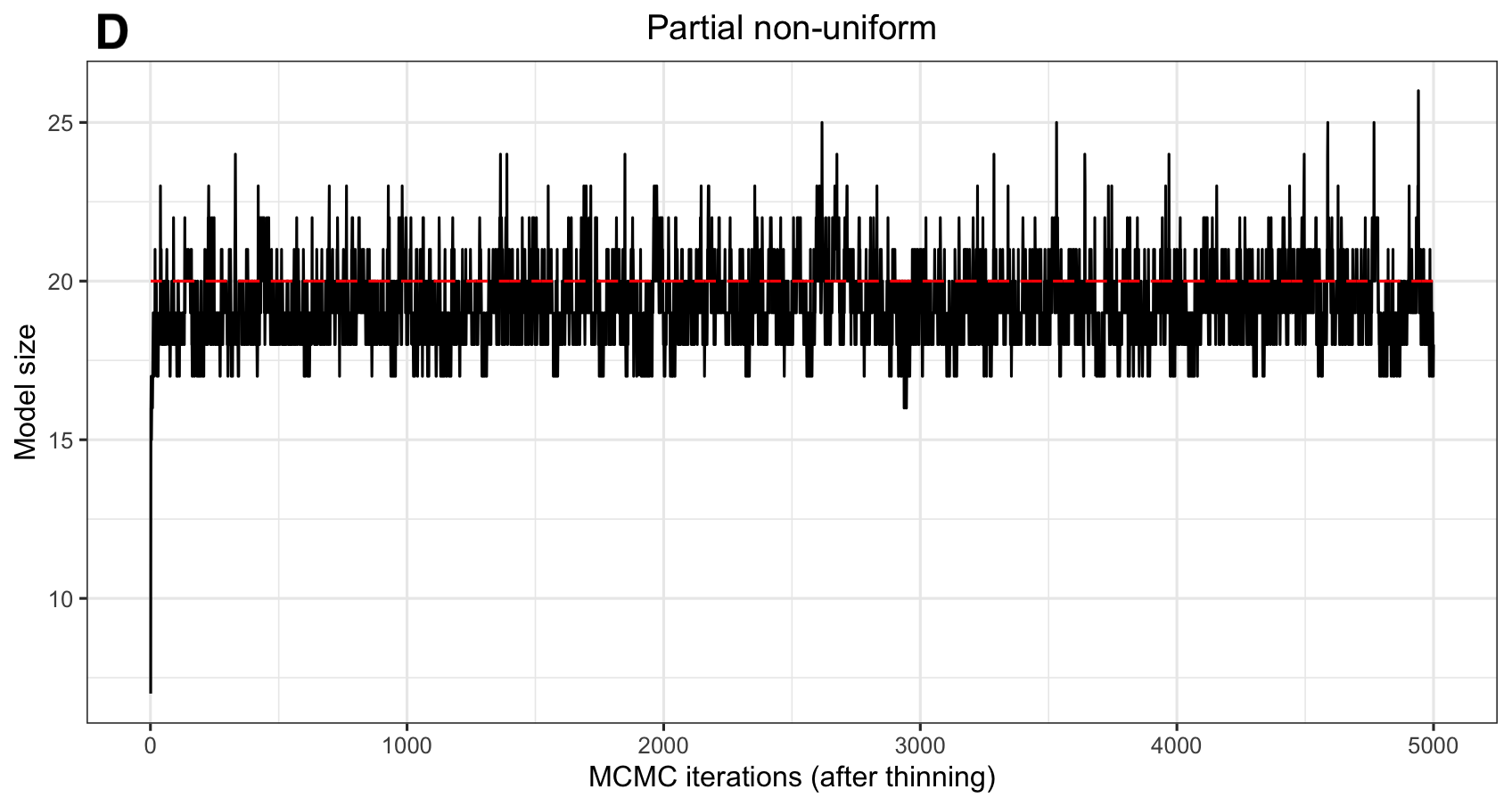}  
\end{minipage}

\begin{minipage}{0.5\textwidth}
\includegraphics[width=0.9\textwidth]{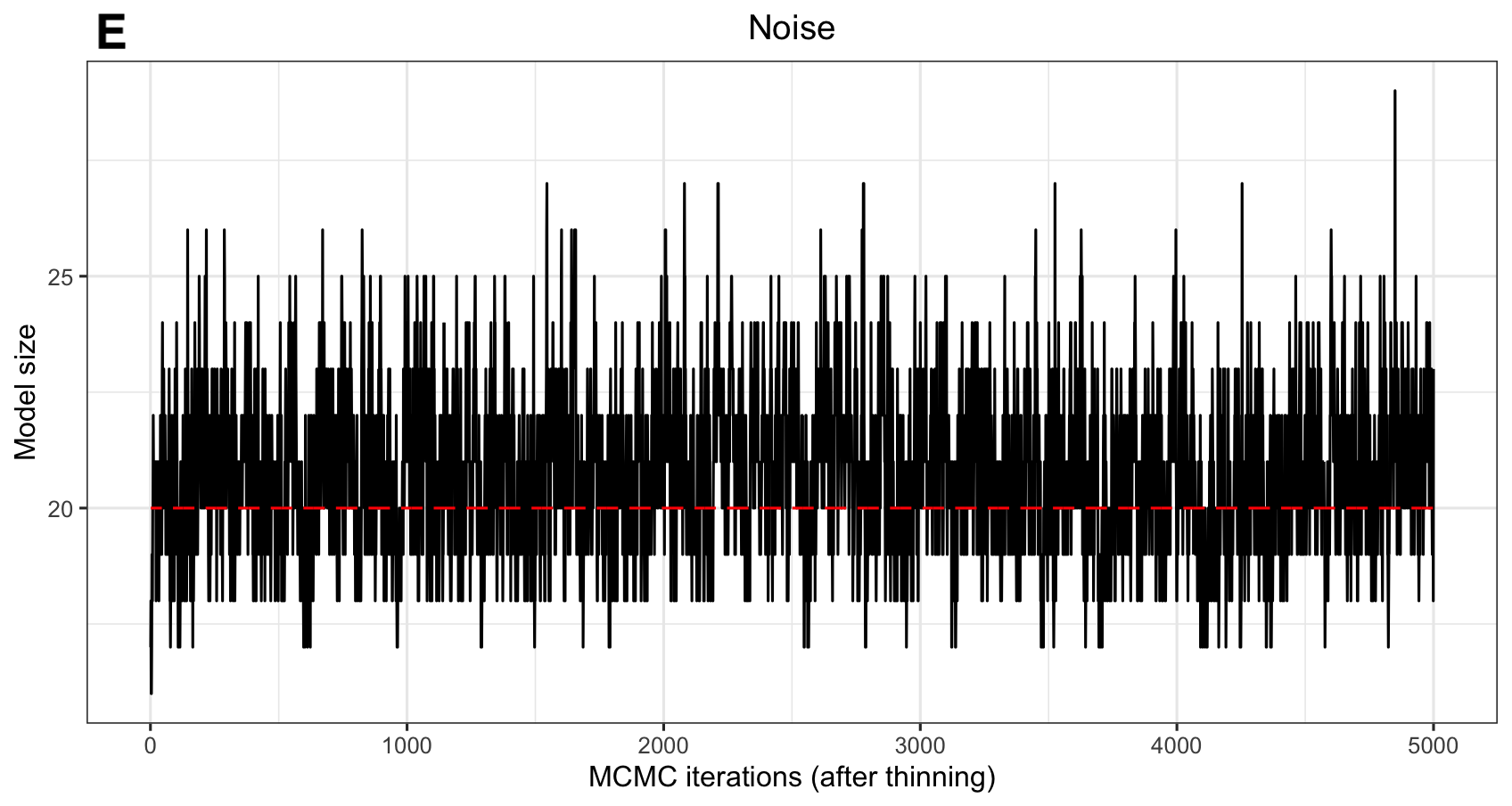}    
\end{minipage}%
\hfill
\begin{minipage}{1.0\textwidth}
    \caption{Results for Simulation study I: Model size for each MCMC iteration after thinning (with thinning parameter $6$) for one of the data sets. Model size is defined as the number of posterior mean inclusion probabilities greater than $0.5$. The red dotted line marks the true model size, which is $20$.}
    \label{fig:mcmc-model-size}
\end{minipage}%
\end{figure}

\begin{figure}[h!tbp]
    \centering
    \textbf{Log-likelihood plots for MCMC diagnostics}\par\medskip

\begin{minipage}{0.5\textwidth}
    \includegraphics[width=0.9\textwidth]{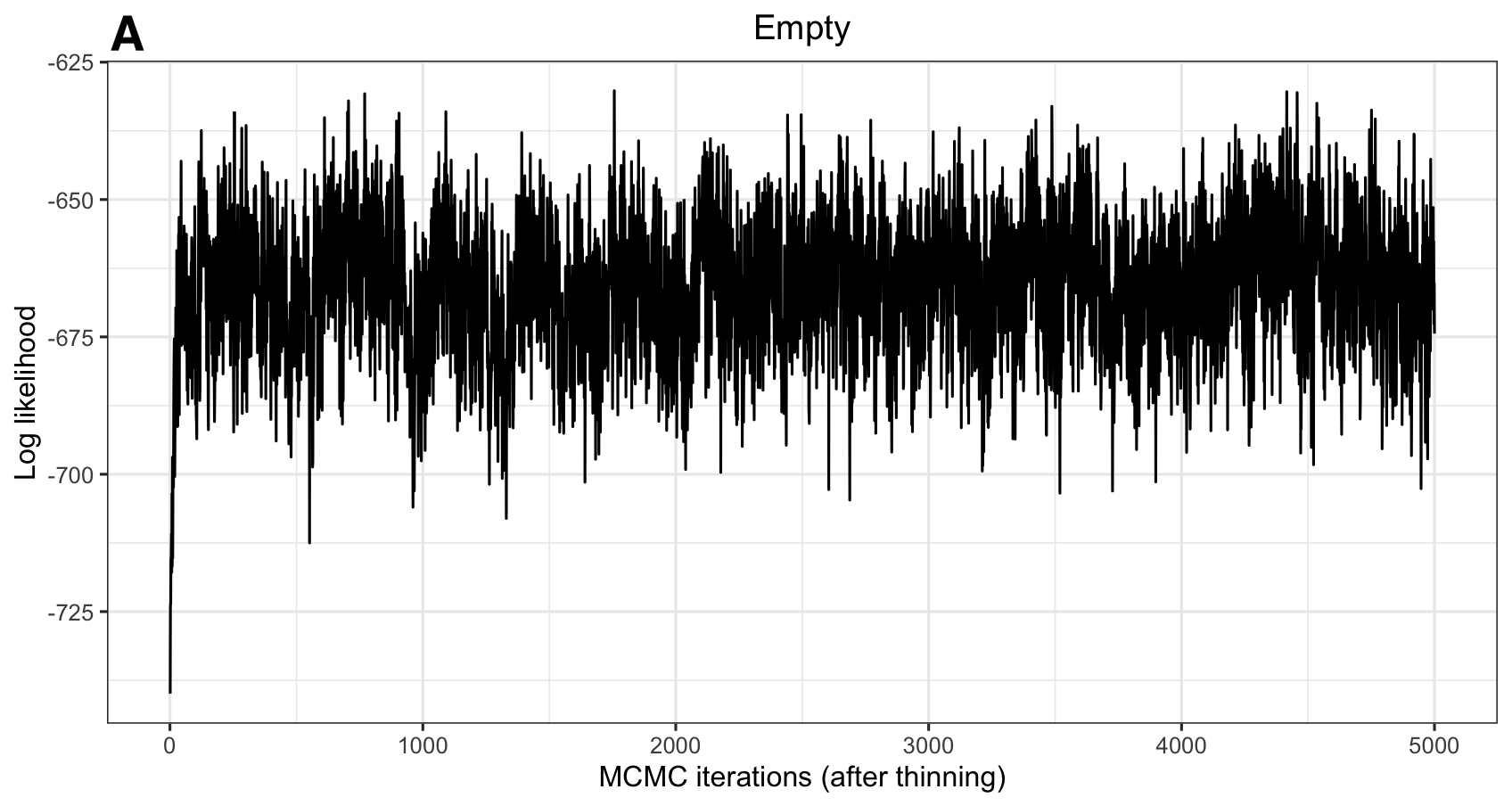}
\end{minipage}%
\hfill
\begin{minipage}{0.5\textwidth}
    \includegraphics[width=0.9\textwidth]{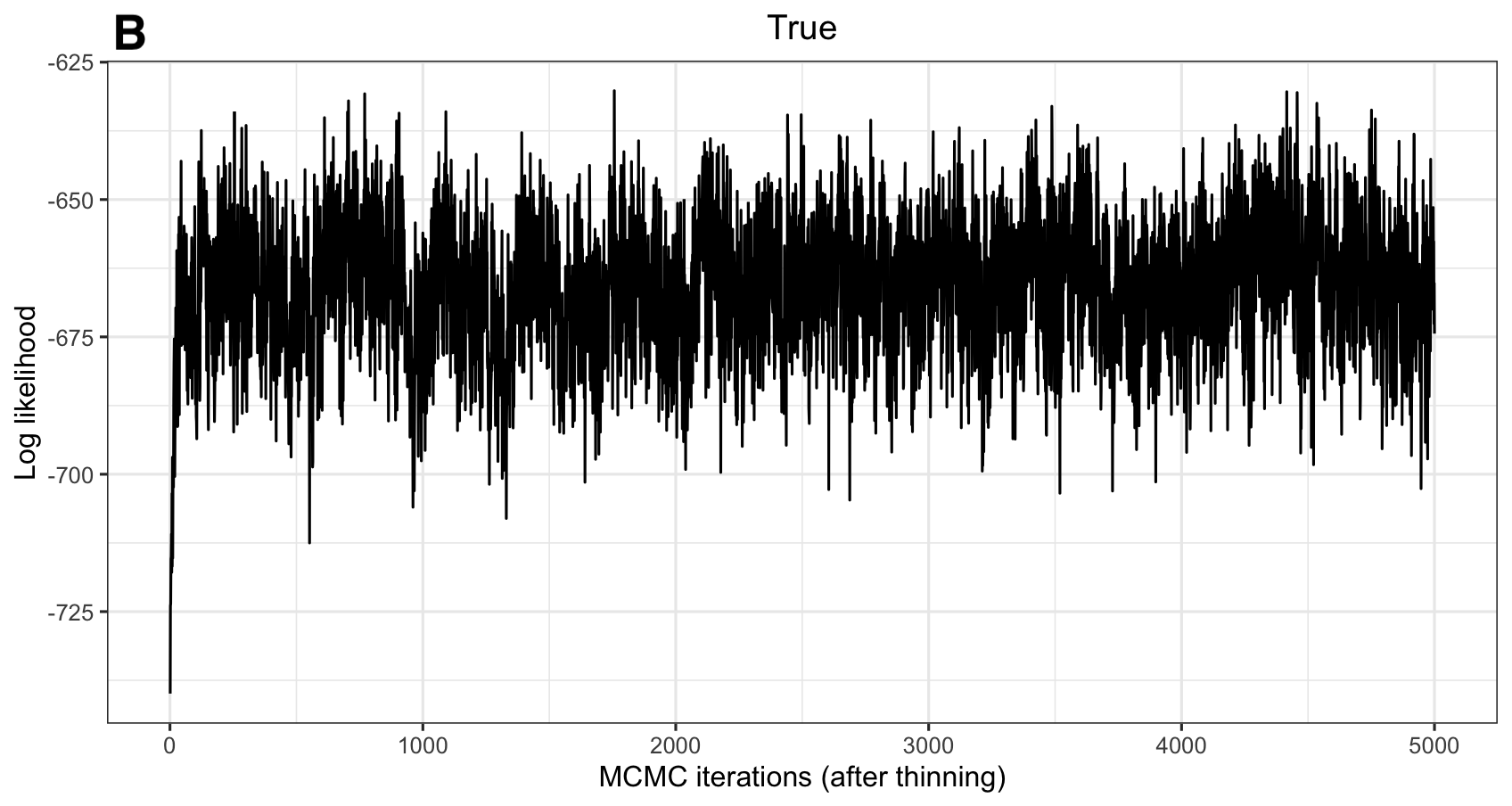}   
\end{minipage}

\begin{minipage}{0.5\textwidth}
    \includegraphics[width=0.9\textwidth]{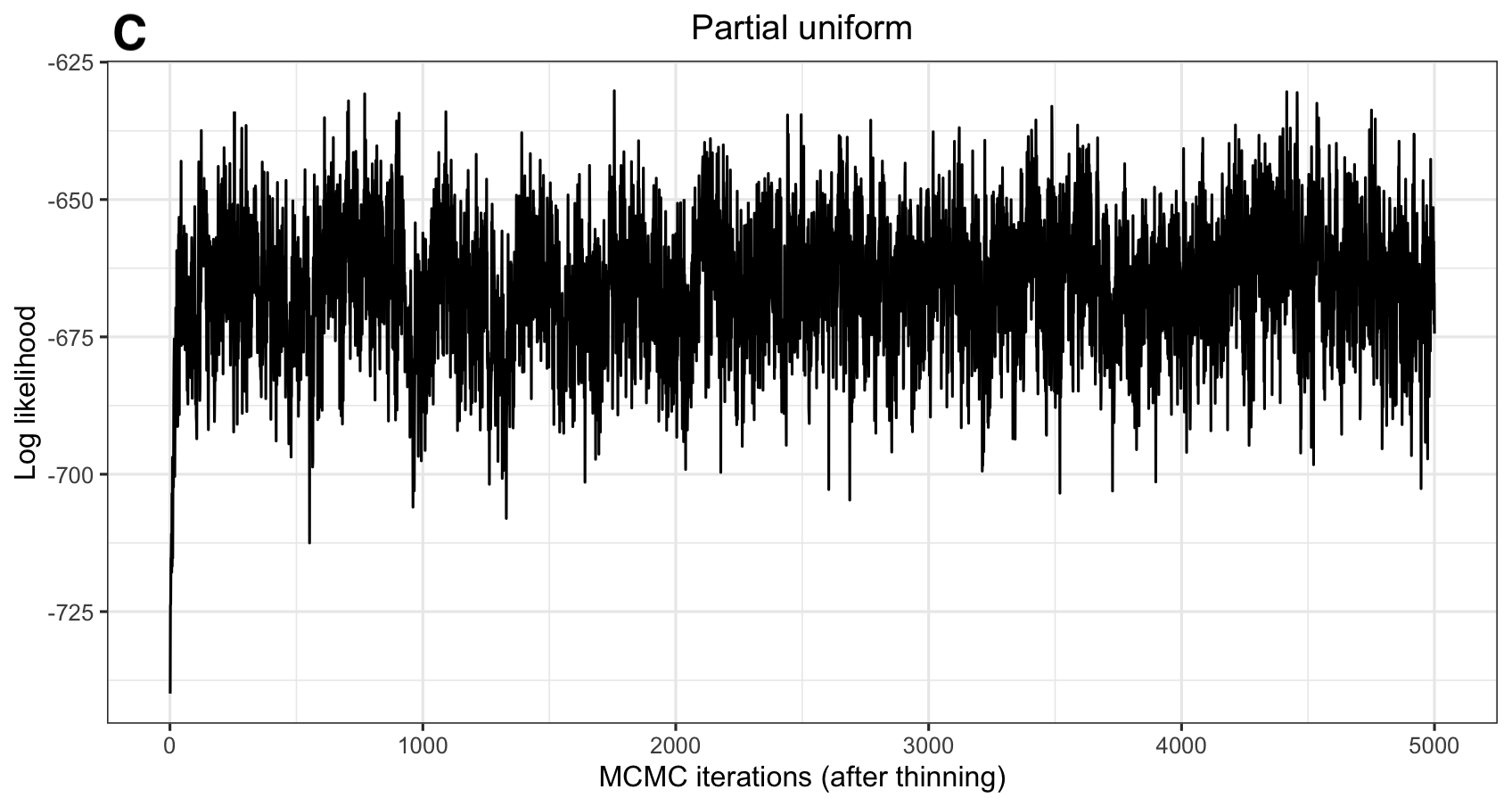} 
\end{minipage}%
\hfill
\begin{minipage}{0.5\textwidth}
    \includegraphics[width=0.9\textwidth]{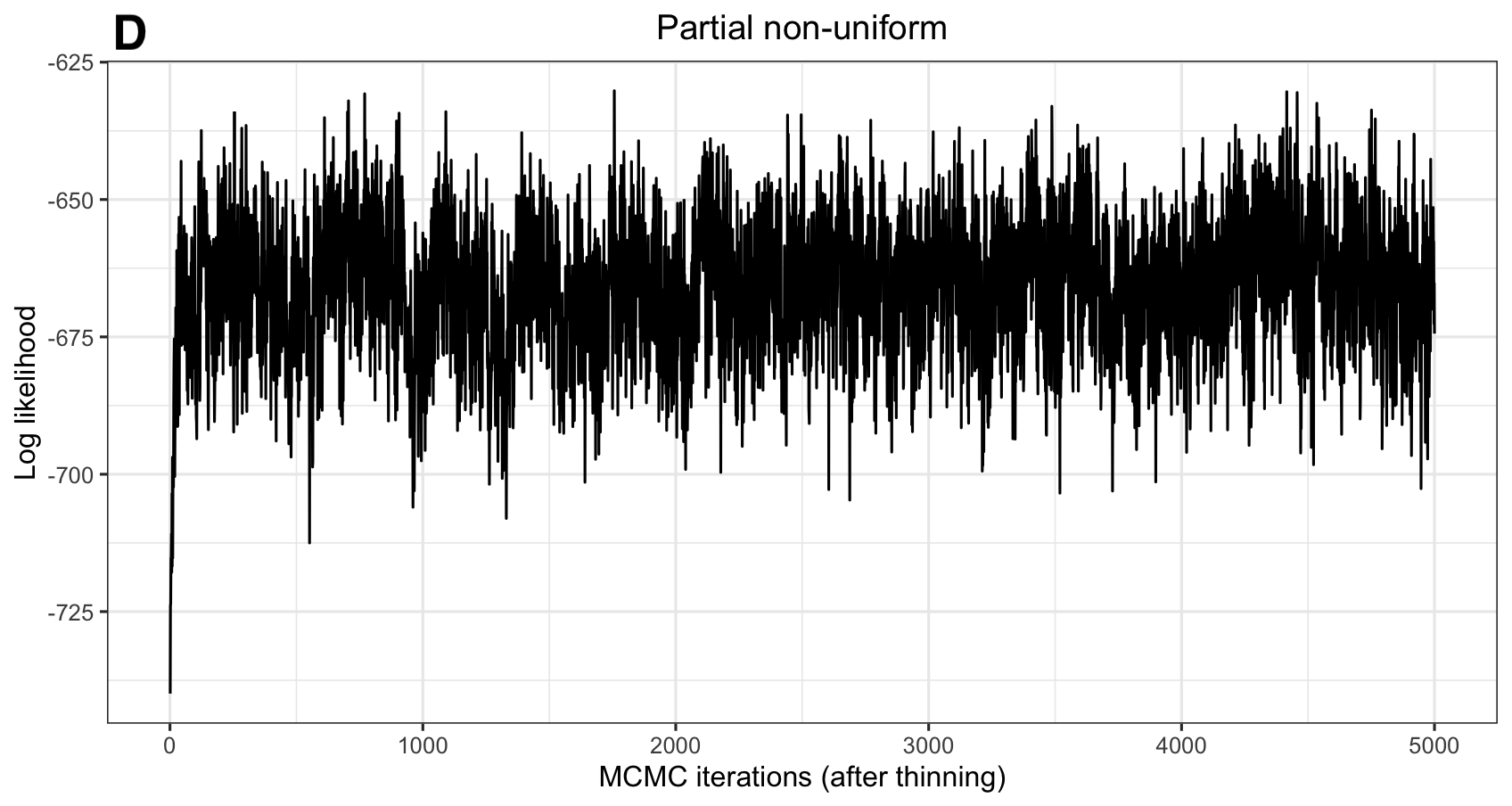} 
\end{minipage}

\begin{minipage}{0.5\textwidth}
\includegraphics[width=0.9\textwidth]{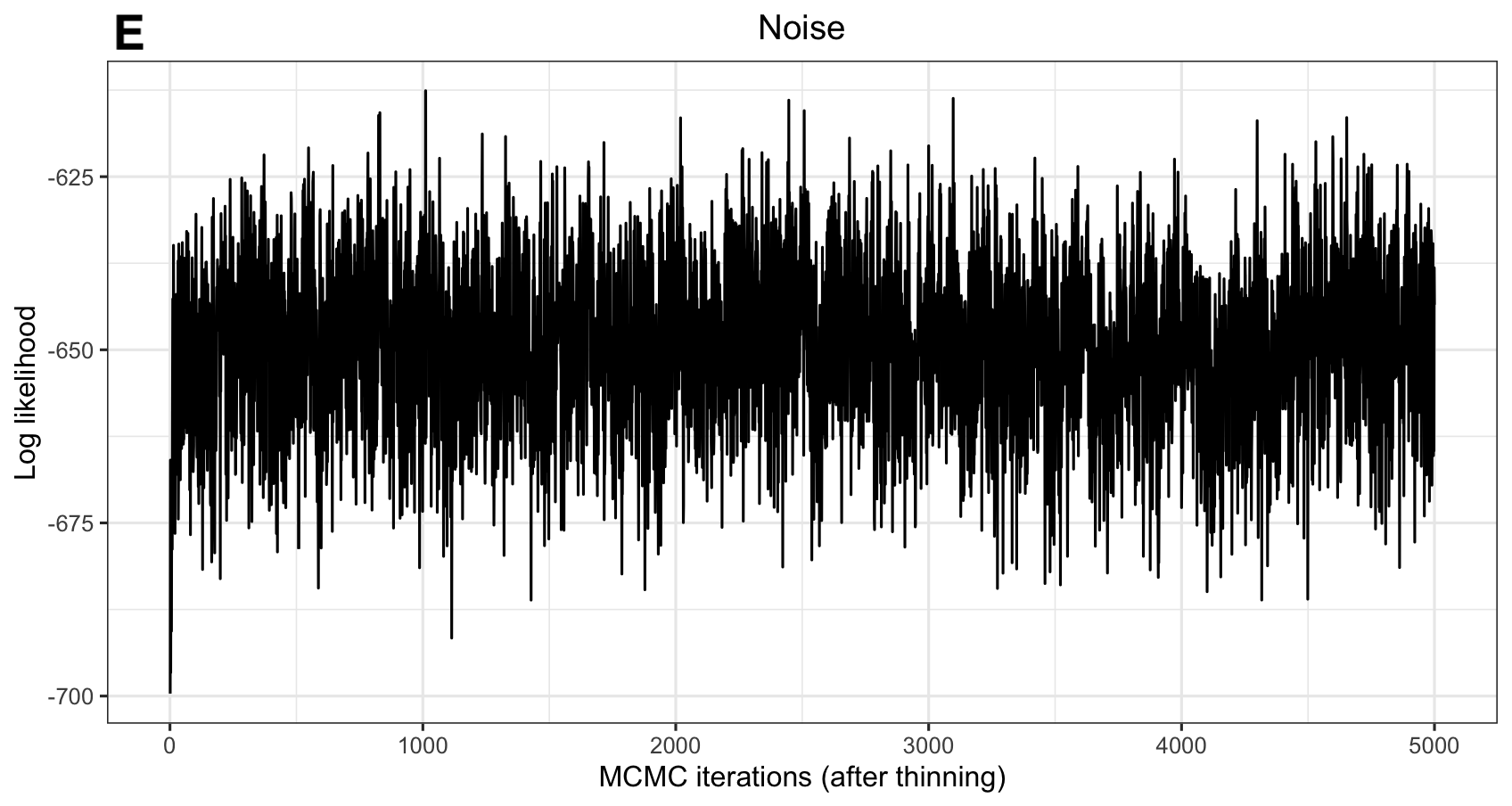} 
\end{minipage}%
\hfill
\begin{minipage}{1.0\textwidth}
    \caption{Results for Simulation study I: The log-likelihood value for the model for each MCMC iteration after thinning (with thinning parameter $6$) for data set 1. }
    \label{fig:mcmc-loglik}
\end{minipage}%
\end{figure}

\begin{figure}

\begin{minipage}{0.5\textwidth}
    \includegraphics[width=0.9\textwidth]{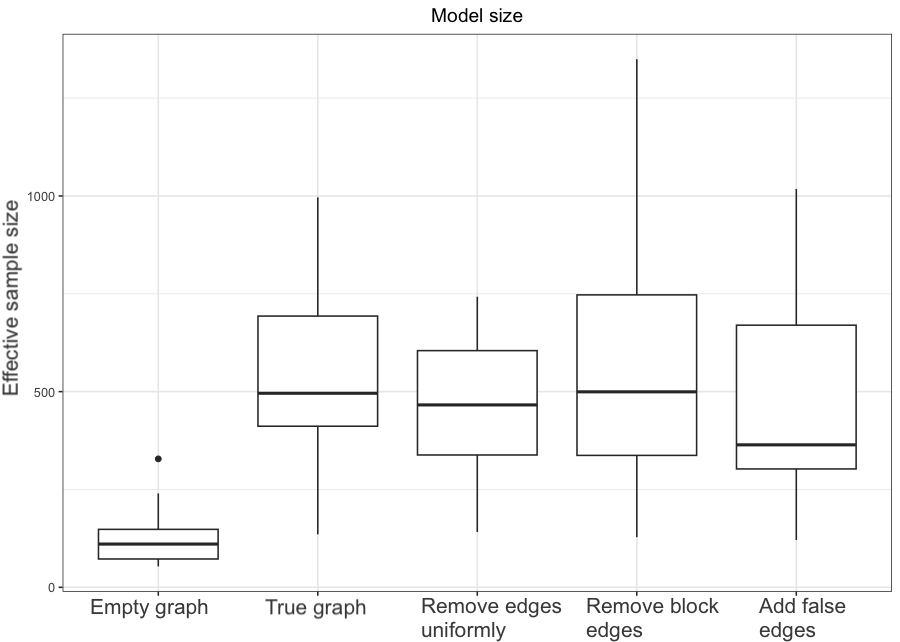} 
\end{minipage}
\hfill
\begin{minipage}{0.5\textwidth}
\includegraphics[width=0.9\textwidth]{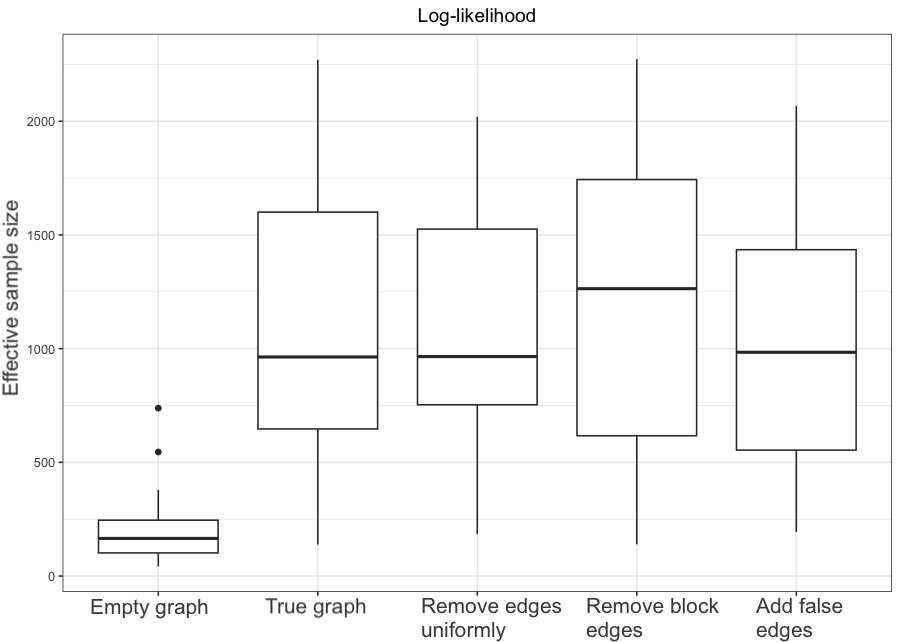}
\end{minipage}%
\hfill
\begin{minipage}{1.0\textwidth}
    \caption{Results for Simulation study I: Effective sample size of the MCMC samples for model size and log-likelihood, for all 20 training data sets. This is based on the MCMC samples after thinning and warmup, which is $2500$ MCMC samples.}
    \label{fig:sim:eff_sample_size}
\end{minipage}%
\end{figure}

\begin{figure}[!htbp]
    \centering
    \includegraphics[width=0.5\textwidth]{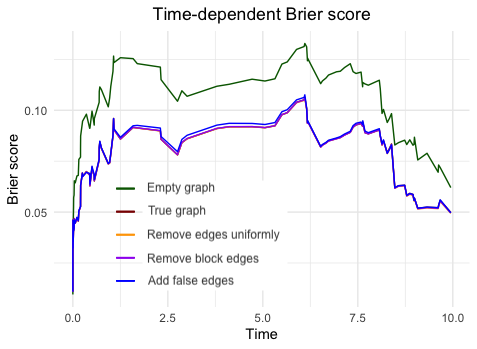}
    \caption{Results for Simulation study I: Time-dependent Brier score on the test set, for one of the $20$ training data sets.}
    \label{fig:time-dep-brier}
\end{figure}

\begin{figure}[h!tbp]
\centering
\includegraphics[width=0.6\textwidth]{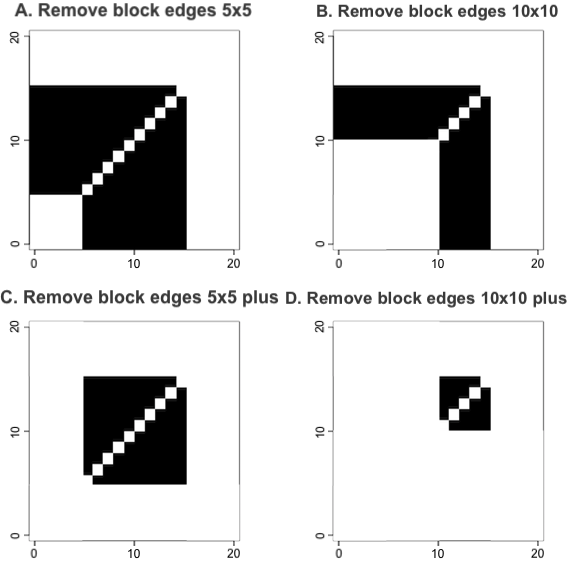}
\caption{Simulation scenario for Simulation study II: The adjacency matrices of the four different graphs used in the MRF prior in Simulation study II. 
Black entries indicate values of 1 and white entries indicate values of 0. 
Note that the plots only show the part of the adjacency matrix corresponding to the first $20$ truly relevant covariates, as all other entries are $0$.}
\label{fig:mrf_graphsII}
\end{figure}

\begin{figure}[!htbp]
    \centering
    \includegraphics[width = 0.7\textwidth]{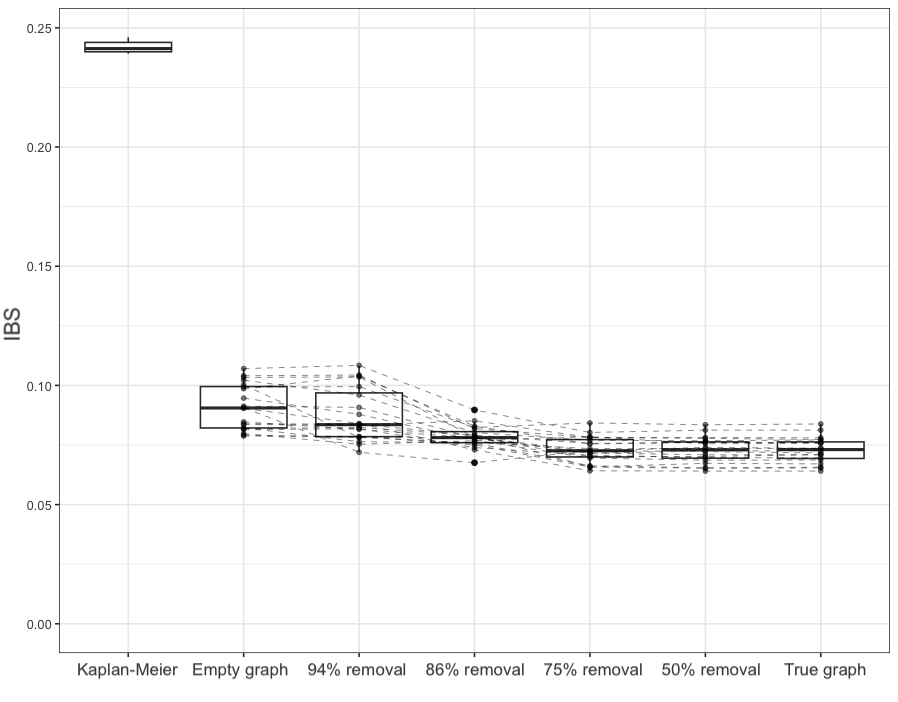}


    \caption{Results for Simulation study II: IBS on a test set based on the MPMs for the Bayesian Cox models with remove edges uniformly to different extents. Dotted lines are connecting the runs for the same data sets.}

    \label{fig:uniform_ibs}
\end{figure}

\begin{figure}
    \centering
    \includegraphics[width = 0.7\textwidth]{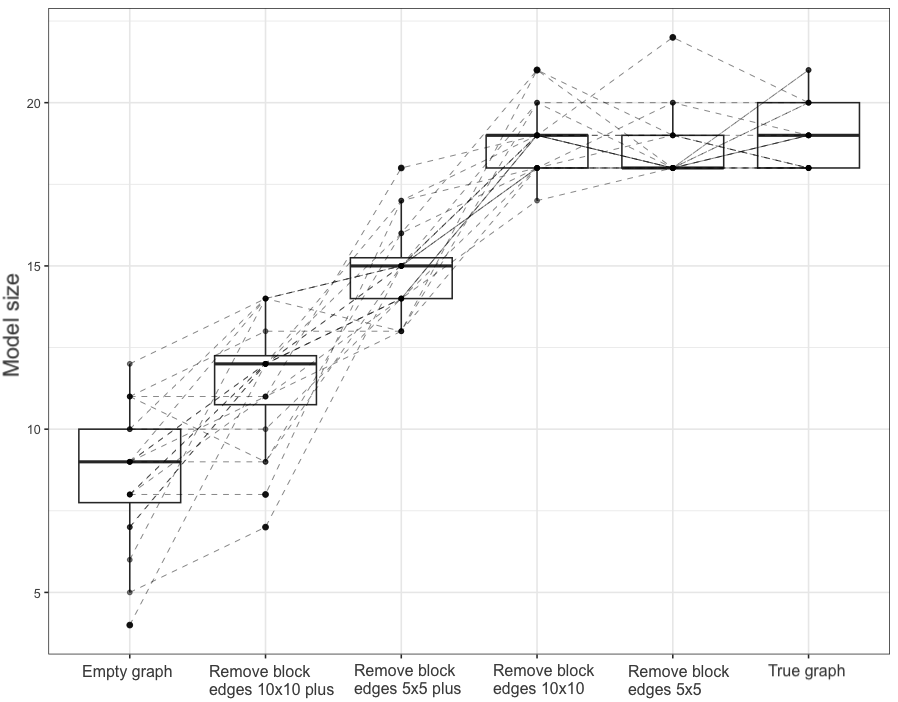}
    \caption{Results for Simulation study II: Model size of the MPMs with removing block edges. Each dot is the model size of the resulting MPM for each training set, and the dotted lines connect the model sizes belonging to the same training set across different models.}
    \label{fig:non-uniform_model_size}
\end{figure}

\begin{figure}[!htbp]
    \centering
    \includegraphics[width = 0.7\textwidth]{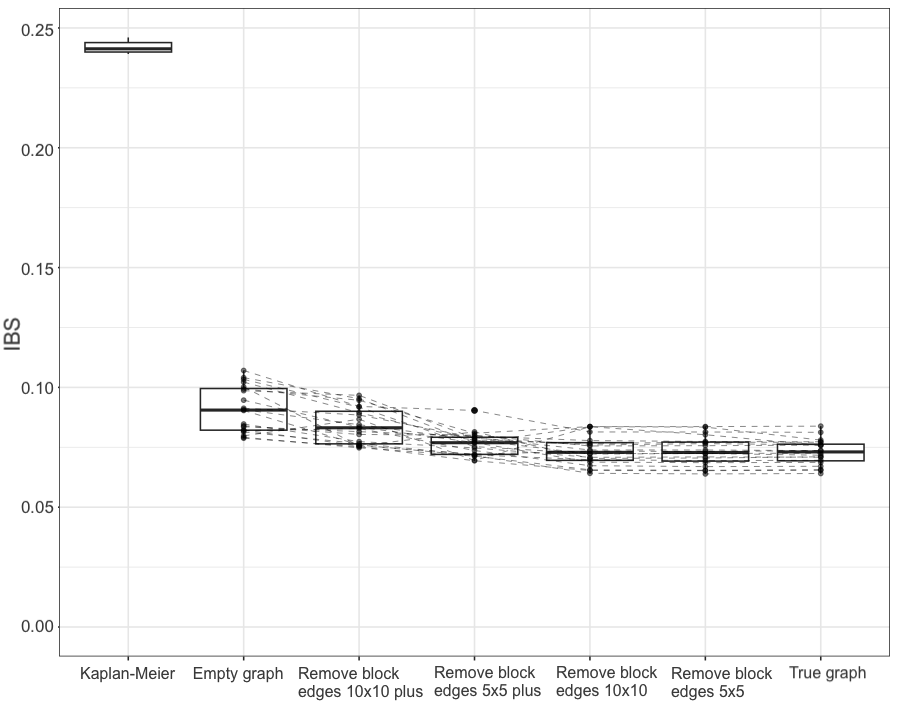}


    \caption{Results for Simulation study II: IBS on a test set based on the MPMs for the Bayesian Cox models with graphs by removing block edges to different extents. Dotted lines are connecting the runs for the same data sets. }
    
    \label{fig:non-uniform_ibs}
\end{figure}

\begin{figure}[!ht]
    \centering
    \includegraphics[width = 0.7\textwidth]{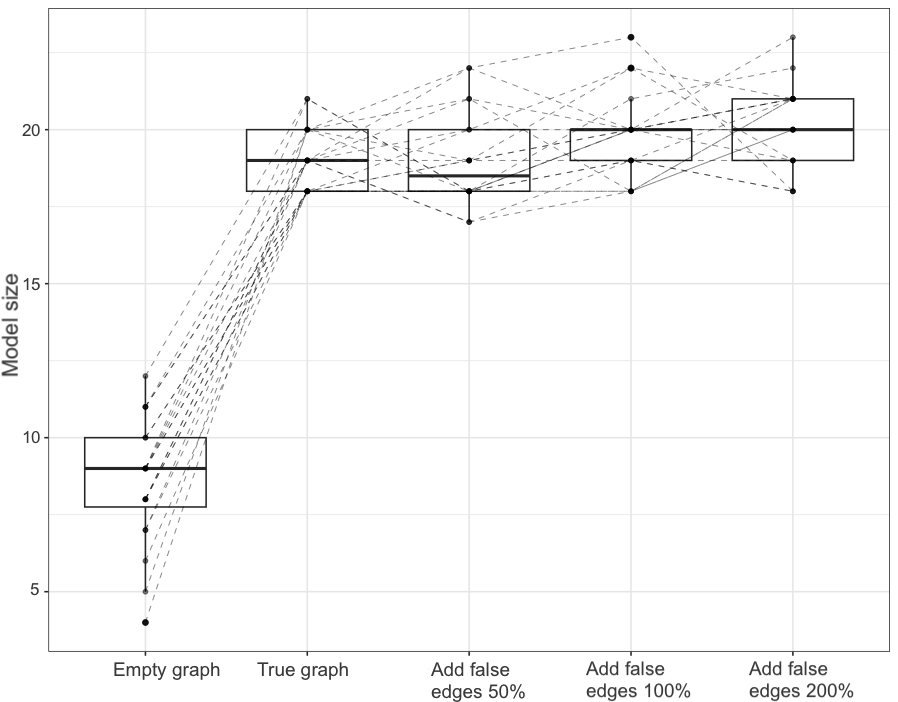}
    \caption{Results for Simulation study II: Model size of the MPM of noisy graph models}
    \label{fig:noise_model_size}
\end{figure}

\begin{figure}[!ht]
    \centering
    \includegraphics[width=0.5\textwidth]{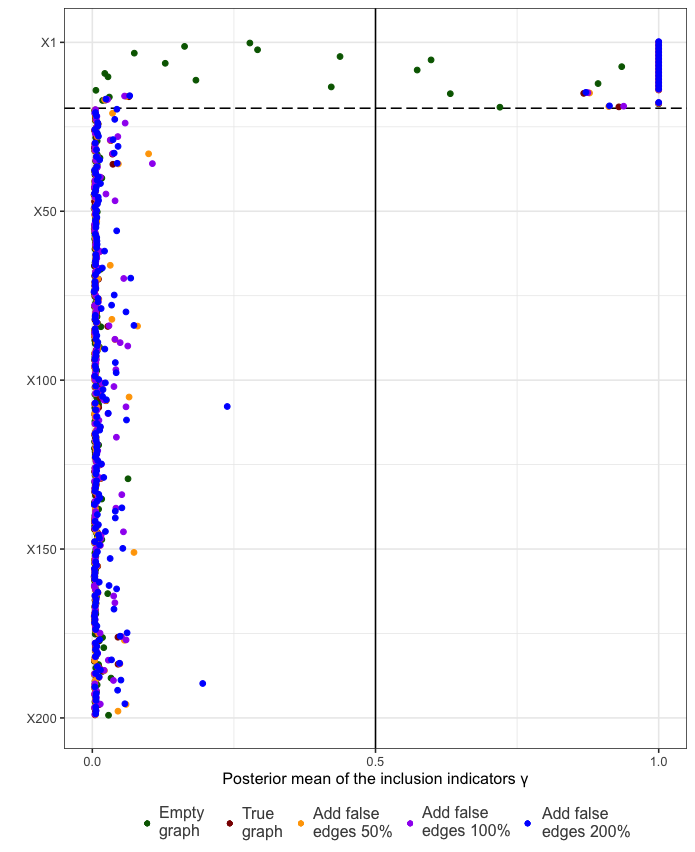}
    \caption{Results for Simulation study II: Mean posterior inclusion probability of all $200$ covariates. The dotted horrizontal line indicates the border between the truly relevant covariates ($1$ to $20$) and the irrelevant covariates.}
    \label{fig:noise_manhattan}
\end{figure}

\begin{figure}
    \centering
    \includegraphics[width =0.7 \textwidth]{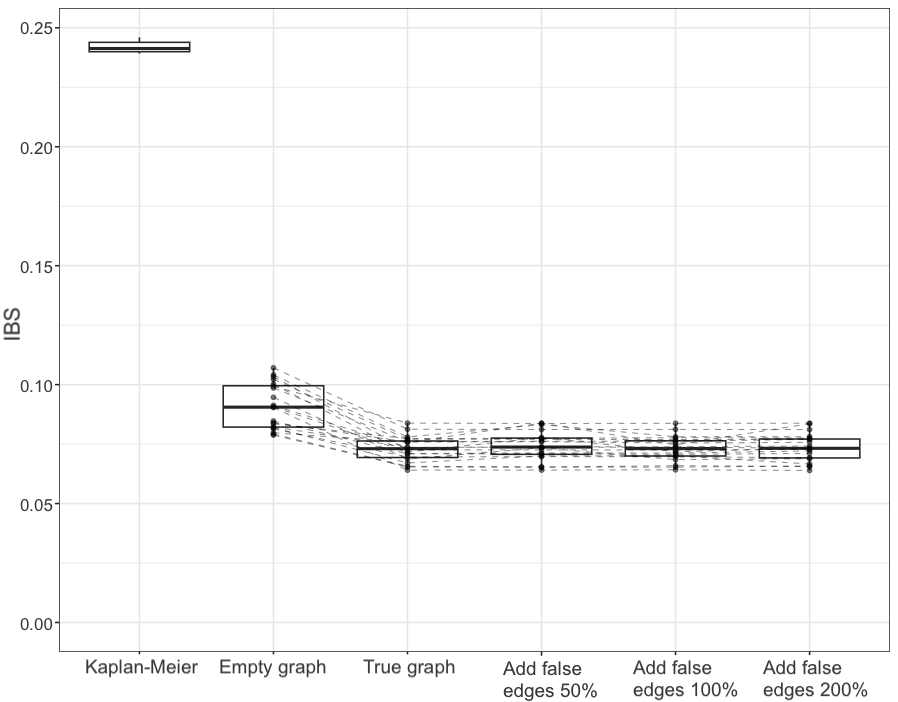}

    
    \caption{Results for Simulation study II: Integrated Brier score of the different levels of noisy graph models, compared to the empty and true graph models. The IBS of the Kaplan-Meier estimates are also included for reference. 
    }
    \label{fig:noise_ibs}
\end{figure}

\clearpage

\begin{landscape}

{\bf S3. Additional results for TCGA data analysis}

\vspace{-3mm}
\begin{figure}[!htbp]
    \centering
    
    \makebox[\textwidth][c]{
    \includegraphics[width=0.91\linewidth]{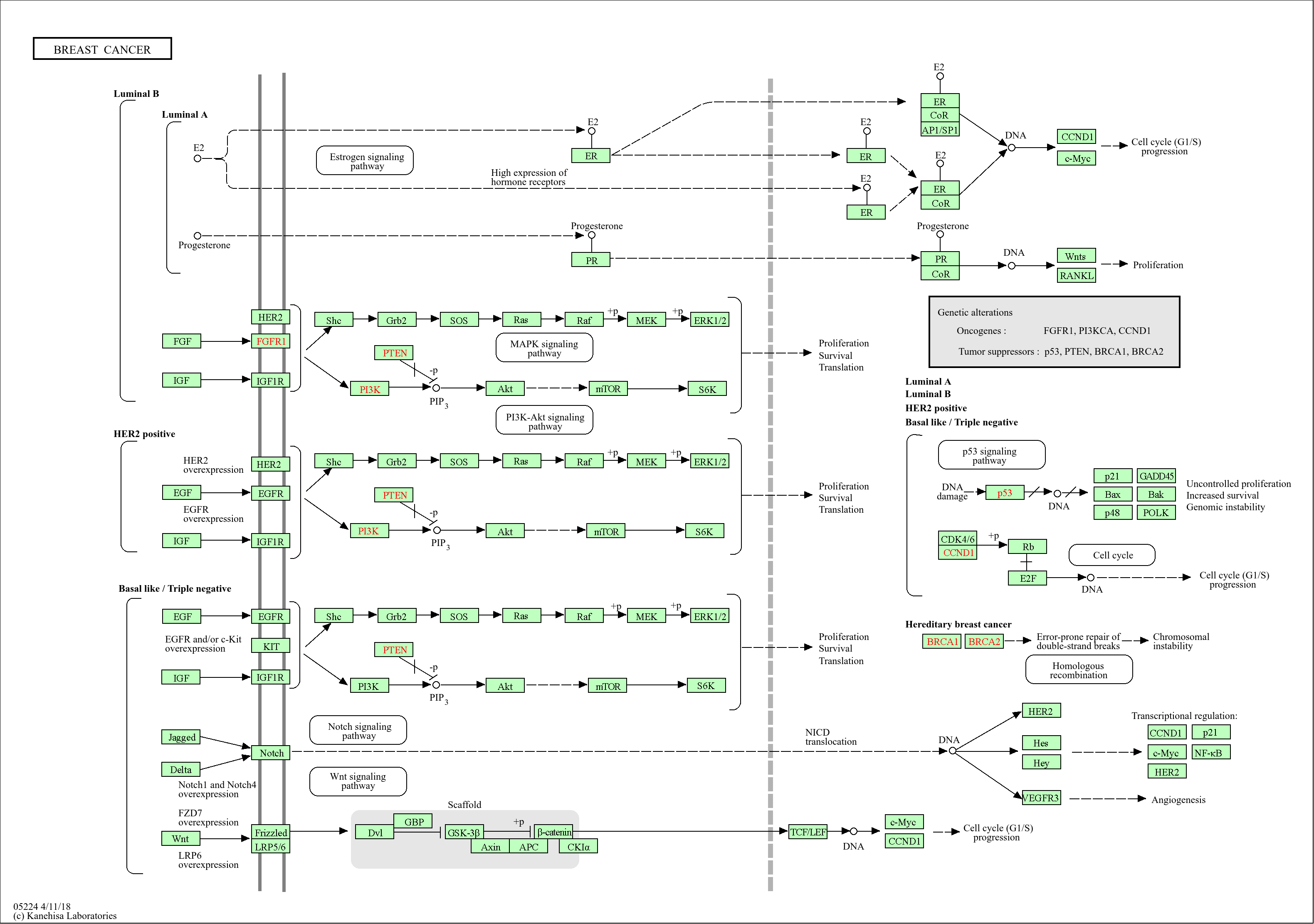}
}
\caption{TCGA data application: The entire networks of KEGG pathways related to breast cancer. Downloaded from \url{https://www.genome.jp/pathway/hsa05224} (April 18, 2024, date last accessed).}
    
    \label{fig:kegg_brca}
\end{figure}
\end{landscape}

\begin{figure}
    \centering
    \includegraphics[width=0.5\linewidth]{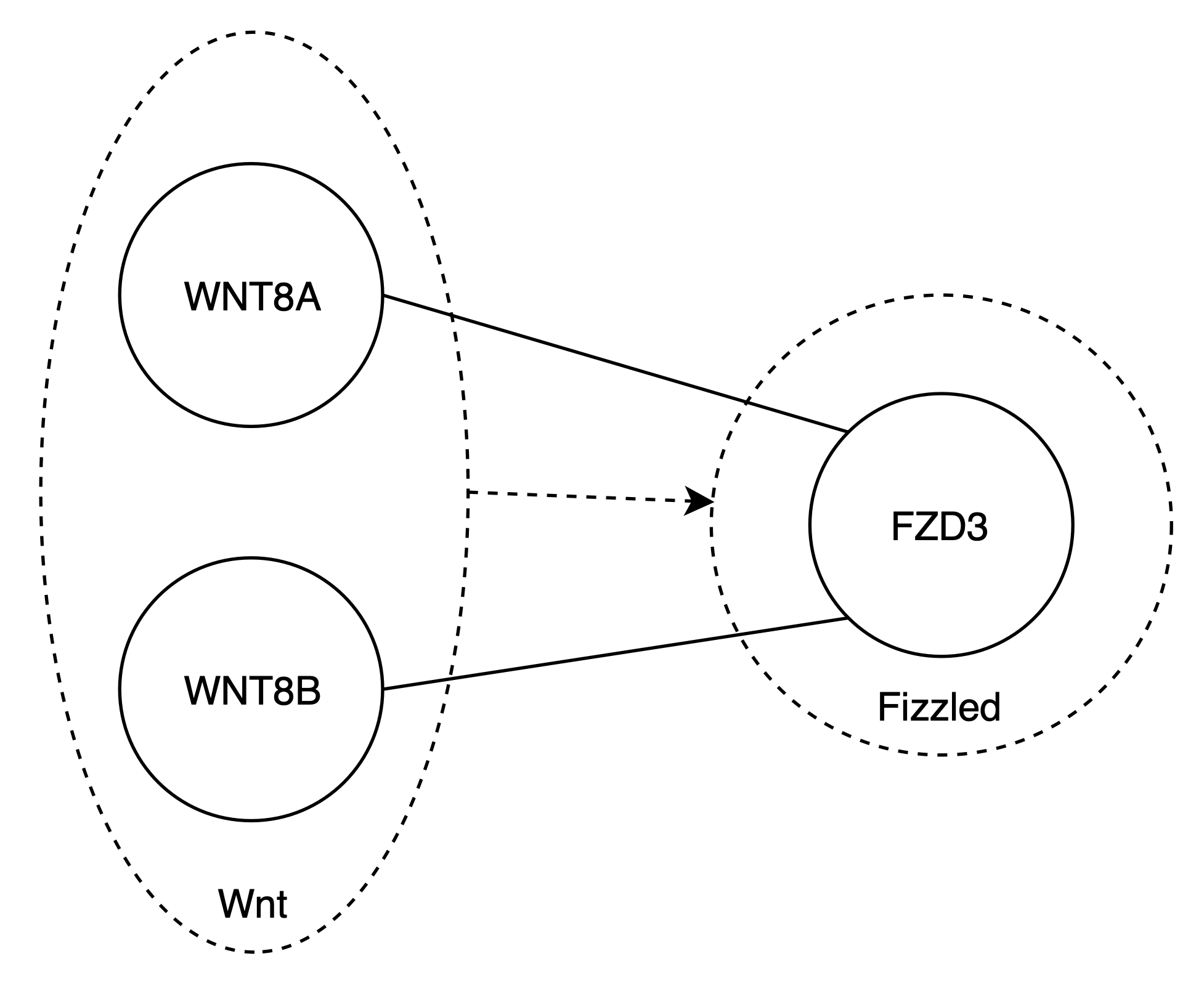}
    \caption{TCGA data application: Illustration of how the graph is constructed from having proteins as nodes (the dotted lines) in the KEGG LRP6-overexpression to Wnt signaling pathway (Wnt $\rightarrow$ Fizzled), to having gene covariates as nodes (solid lines) in the graph used in the MRF prior. Note that while only two genes are chosen as representatives here (WNT8A and WNT8B), our data set actually includes a total of 18 gene expression features (and no mutations), see Figure \ref{fig:kegg_in_graph}. The edges are also changed from being directed in the KEGG pathways to being undirected in the graph for the MRF prior. }
    \label{fig:graph_construction}
\end{figure}

\begin{table}[!htbp]
    \centering
    \caption{TCGA data application: The number of edges and the number of nodes with at least one neighbour in the graph created from the KEGG breast cancer pathways, for the different types of covariates: gene expression (GEX), mutation (MUT) and clinical. The MUT -- GEX column refers to edges connecting gene expression and mutation features that correspond to the same gene. }
\begin{tabular}{llllll}
\hline
\hline
                                     & GEX & MUT & MUT -- GEX & Clinical & Total \\
                                     \hline

Nodes  & 144             & 21       & -- & 2        & 167  \\
Edges                                & 543             & 70       & 21& 1        & 635   \\
\hline
\hline
\end{tabular}
    \label{tab:graph_composition}
\end{table}

\begin{figure}[!ht]
    \begin{minipage}{0.5\textwidth}
        \includegraphics[width=0.9\textwidth]{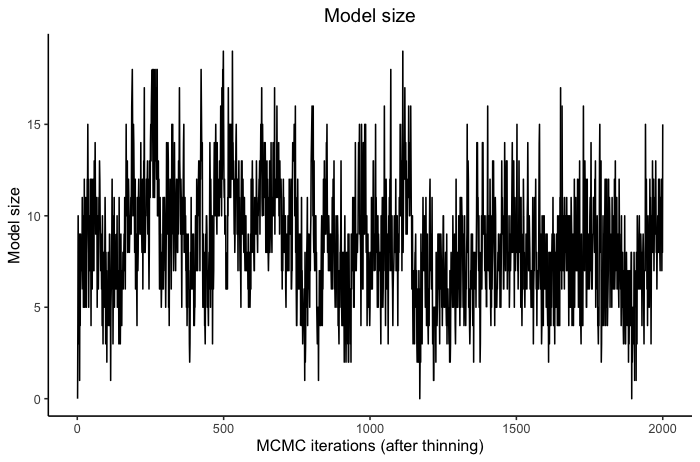}
    \end{minipage}
    \hfill
    \begin{minipage}{0.5\textwidth}
    \includegraphics[width=0.9\textwidth]{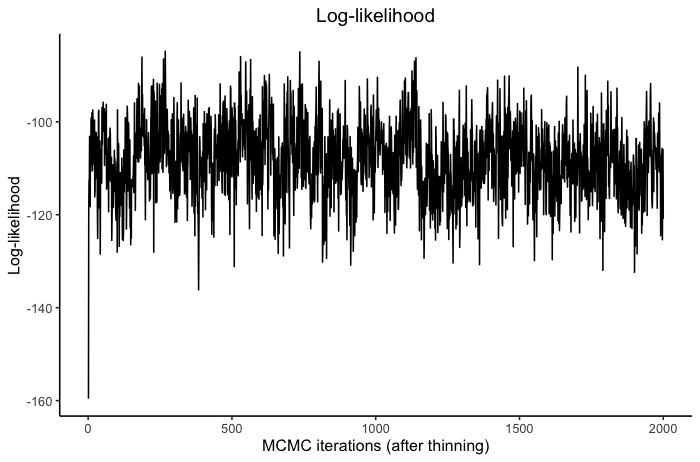}
    \end{minipage}%
    \hfill
    \begin{minipage}{1.0\textwidth}
        \caption{TCGA data application: Plots of the model size and the log-likelihood for each iteration after thinning for the empty graph model, for one of the data splits.}
    \label{fig:mcmc_empty_real}
    \end{minipage}%
\end{figure}
\clearpage

\begin{figure}[!ht]
    \begin{minipage}{0.5\textwidth}
        \includegraphics[width=0.9\textwidth]{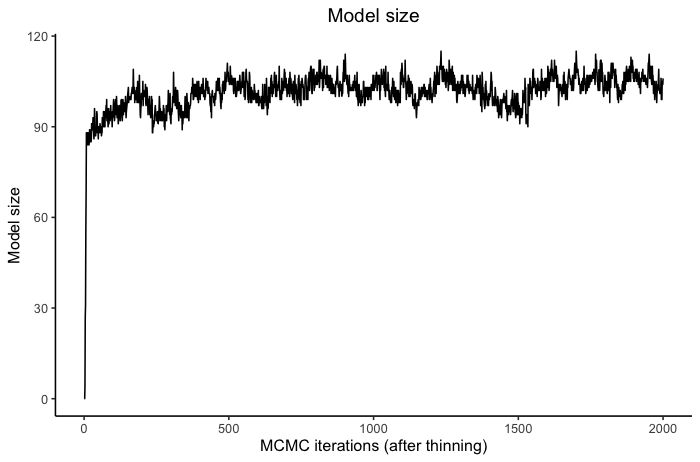}
    \end{minipage}
    \hfill
    \begin{minipage}{0.5\textwidth}
    \includegraphics[width=0.9\textwidth]{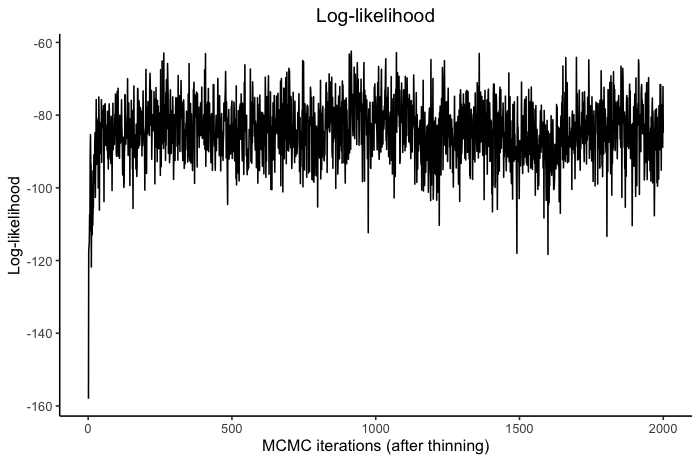}
    \end{minipage}%
    \hfill
    \begin{minipage}{1.0\textwidth}
    \caption{TCGA data application: Plots of the model size and the log-likelihood for each iteration after thinning for the KEGG graph model, for one of the data splits.}
    \label{fig:mcmc_kegg}
    \end{minipage}%
\end{figure}

\begin{figure}[!ht]
    \begin{minipage}{0.5\textwidth}
        \includegraphics[width=0.9\textwidth]{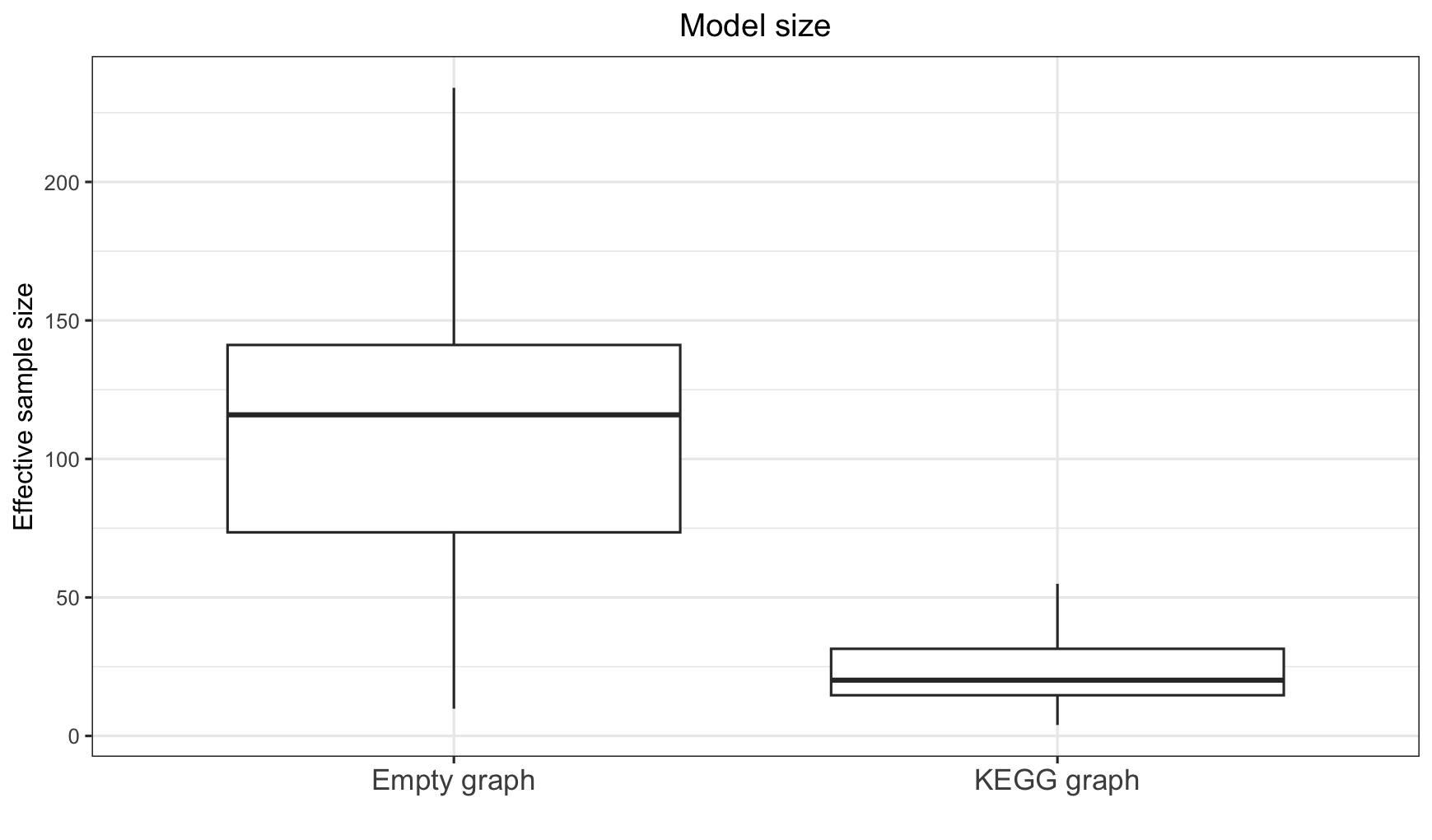}
    \end{minipage}
    \hfill
    \begin{minipage}{0.5\textwidth}
    \includegraphics[width=0.9\textwidth]{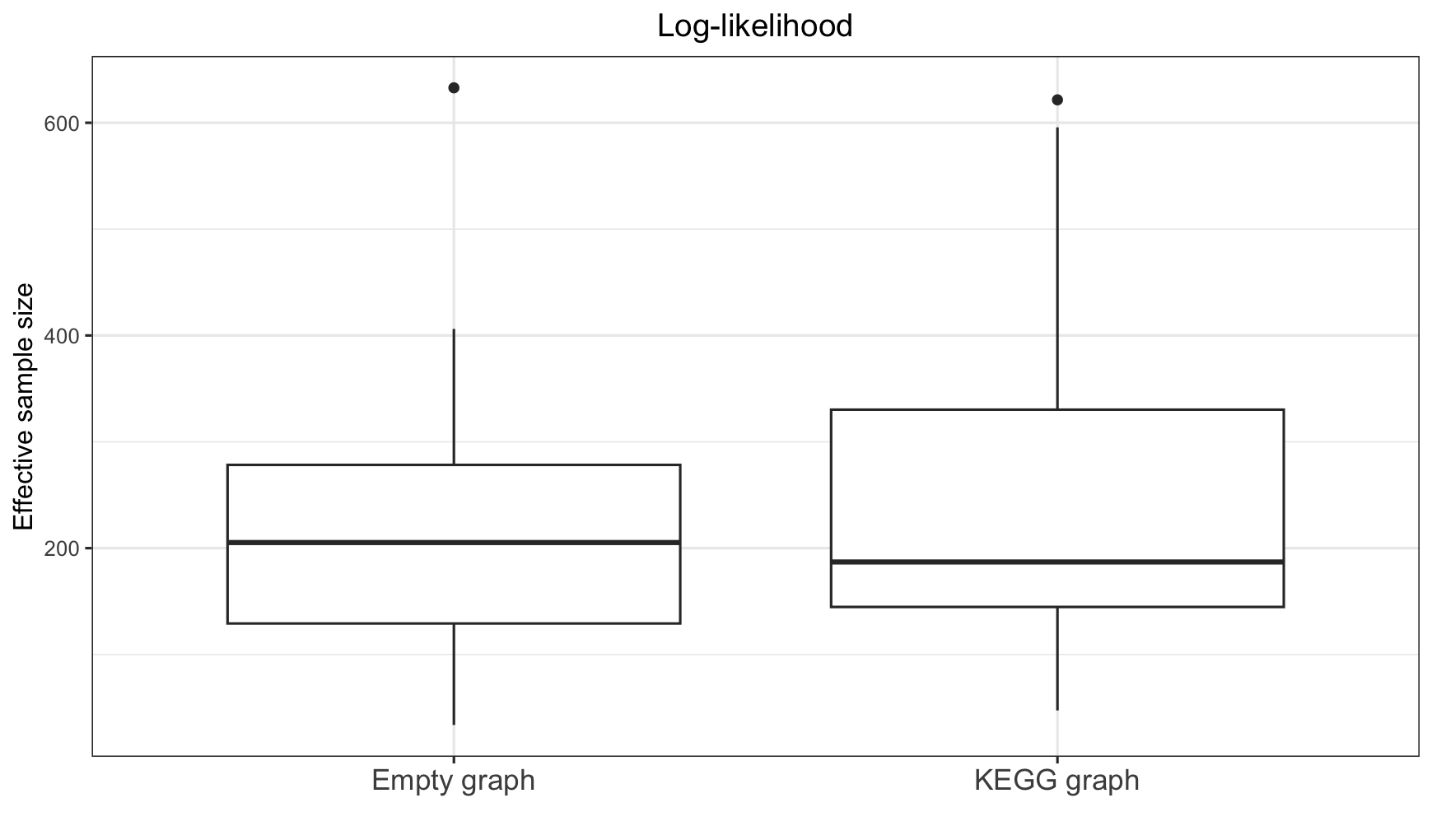}
    \end{minipage}%
    \hfill
    \begin{minipage}{1.0\textwidth}
    \caption{TCGA data application: Effective sample size for the model size and the log-likelihood, for the $20$ data splits of both the empty graph model and the KEGG graph model.}
    \label{fig:effective_sample_size}
    \end{minipage}%
\end{figure}

\begin{figure}[!ht]
    \centering
    \includegraphics[width=0.5\linewidth]{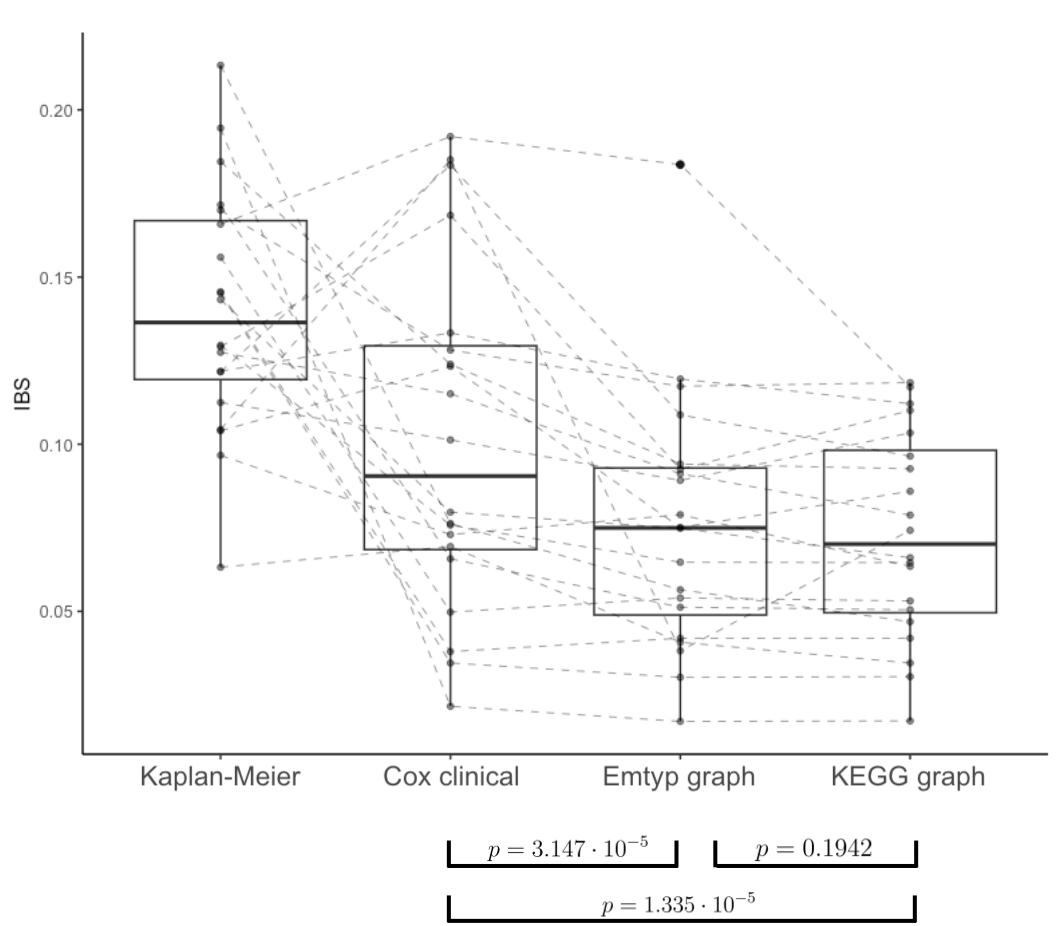}
    \caption{TCGA data application: The integrated Brier score (IBS) of the two Bayesian Cox models, with and without the KEGG graph, compared to the Kaplan-Meier method and the frequentist Cox model with two clinical covariates age and treatment. The p-values at the bottom of the boxplot are for the one-sided Wilcoxon signed rank test, with the alternative hypothesis that the model to the right has smaller IBS than the model to the left. }
    \label{fig:ibs_real}
\end{figure}


\begin{figure}[!htbp]
    \centering
    \includegraphics[width=0.5\linewidth]{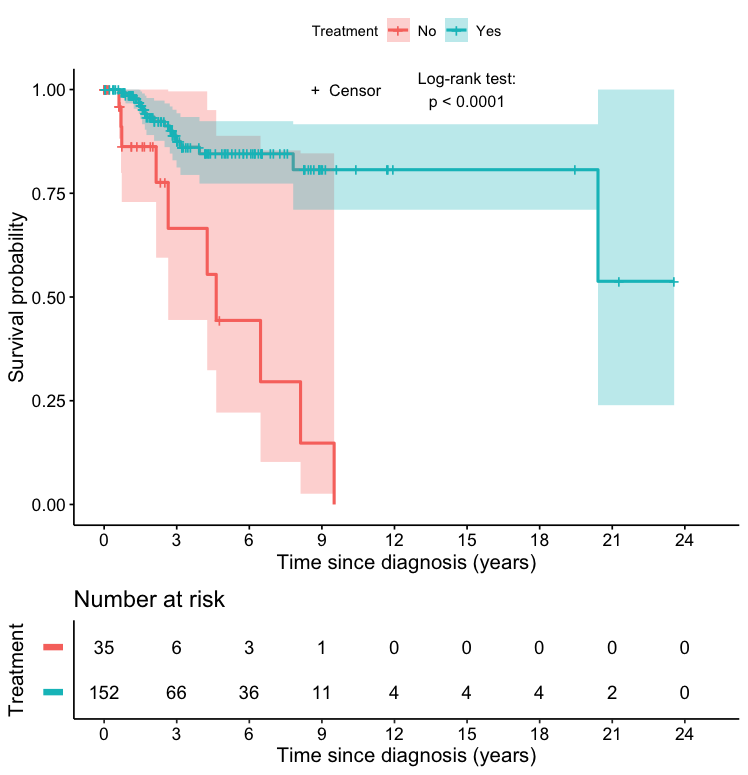}
    \caption{TCGA data application: The Kaplan-Meier estimates of the survival function for the basal subtype of the TCGA breast cancer data set, stratified based on whether or not the patient received treatment (i.e. pharmaceutical/radiation therapy or not).}
    \label{fig:tcga_basal}
\end{figure}

\clearpage

\end{document}